\documentclass[12pt]{article}
\pdfoutput=1
\usepackage{jheppub}
\usepackage{amsmath}
\usepackage{amssymb}
\usepackage{wrapfig}
\usepackage{bbm}
\usepackage{arydshln}
\usepackage{multirow}
\usepackage{mathtools}
\usepackage{blkarray}
\usepackage{setspace}
\usepackage{dsfont}
\usepackage{float}
\usepackage{subcaption}
\usepackage{braket}
\usepackage[utf8]{inputenc}
\usepackage[english]{babel}
 
\newtheorem{conj}{Conjecture}

\allowdisplaybreaks

\title{On Miura Transformation for $\mathcal{W}_{m|n\times \infty}$ and Calabi-Yau Singularities}
\title{On Extensions of $\widehat{\mathfrak{gl}(m|n)}$ Kac-Moody algebras and Calabi-Yau Singularities}

\abstract{We discuss a class of vertex operator algebras $\mathcal{W}_{m|n\times \infty}$ generated by a super-matrix of fields for each integral spin $1,2,3,\dots$. The algebras admit a large family of truncations that are in correspondence with holomorphic functions on the Calabi-Yau singularity given by solutions to $xy=z^mw^n$. We propose a free-field realization of such truncations generalizing the Miura transformation for $\mathcal{W}_N$ algebras. Relations in the ring of holomorphic functions lead to bosonization-like relations between different free-field realizations. The discussion provides a concrete example of a non-trivial interplay between vertex operator algebras, algebraic geometry and gauge theory.}

\author[ab]{Miroslav Rap\v{c}\'{a}k}
\affiliation[a]{Perimeter Institute for Theoretical Physics, Waterloo, Canada}
\affiliation[b]{Center for Theoretical Physics, University of California, Berkeley}
\emailAdd{miroslav.rapcak@gmail.com}

\begin{document}
\maketitle

\section{Introduction}

Vertex operator algebras (VOAs) are well known to appear in the context of supersymmetric gauge theories. A prototypical example of such an appearance is the AGT correspondence relating $\mathcal{W}_{N}\times \widehat{\mathfrak{gl}(1)}$ algebras (see e.g. \cite{Zamolodchikov:1985wn,Fateev:1987zh,Drinfeld:1984qv,Bais:1987dc,Goddard:1984vk,Goddard:1986ee,Bais:1987zk,Bowcock:1990ku,Kausch:1990bn,Bershadsky:1989mf,Feigin:1990pn}) and supersymmetric $U(N)$ gauge theories living on $\mathbb{C}^2$ \cite{Alday:2009aq,Wyllard:2009hg}. Concretely, one can identify the Nekrasov partition function of such gauge theories \cite{Nekrasov:2002qd} with conformal blocks of $\mathcal{W}_{N}\times \widehat{\mathfrak{gl}(1)}$ algebras \cite{Alday:2009aq} or construct a natural geometric action of $\mathcal{W}_{N}\times \widehat{\mathfrak{gl}(1)}$ on the equivariant cohomology of the moduli space of ADHM instantons \cite{Nakajima:1994nid,Schiffmann:2012gf,Maulik:2012wi,Braverman:2014xca}.

An interesting perspective on $\mathcal{W}_{N}\times \widehat{\mathfrak{gl}(1)}$ algebras comes from considering the large $N$ limit and treating the parameter $N$ as a generic complex number \cite{Yu:1991bk,deBoer:1993gd,Khesin:1994ey,Hornfeck:1994is,Blumenhagen:1994wg,Gaberdiel:2012aa,Prochazka:2014aa,Linshaw:2017tvv}. The resulting algebra denoted as $\mathcal{W}_{1+\infty}$ is generated by infinitely many fields $U_1,U_2,U_3,\dots$ of spins $1,2,3,\dots$ and depends on two complex parameters\footnote{The algebra $\mathcal{W}_{N}\times \widehat{\mathfrak{gl}(1)}$ already depends on one continuous parameter $\Psi$ related to the central charge of the Virasoro subalgebra.} $N$ and $\Psi$. Furthermore, the algebra $\mathcal{W}_{1+\infty}$ is rigid in the sense that it is the only algebra containing a single strong generator of each integral spin and satisfying Jacobi identities \cite{Gaberdiel:2012aa,Prochazka:2014aa,Linshaw:2017tvv}. Algebras $\mathcal{W}_{N}\times \widehat{\mathfrak{gl}(1)}$ can be then recovered by specializing the parameter $N$ to a positive integer and moding out the ideal generated by fields $U_i$ of spin greater than $N$. 

It was observed in \cite{Prochazka:2017qum} (based on results of \cite{Prochazka:2014aa,Prochazka:2015deb,Bershtein,Litvinov:2016mgi}) that $\mathcal{W}_{1+\infty}$ algebra actually contains a three-parameter family of truncations $Y_{N_1,N_2,N_3}$ parametrized by non-negative integers $N_i$. These more general truncations can be furthermore identified with VOAs from \cite{Gaiotto:2017euk} defined in terms of the quantum Hamiltonian reduction. This cohomological definition is motivated by an analysis of a particular configuration of interfaces in four-dimensional gauge theory that provides us with a physical realization of the algebras. Using well-known dualities of string theory \cite{Leung:1997tw,Nekrasov:2010ka,Prochazka:2017qum,Rapcak:2018nsl}, one can relate the setup of \cite{Gaiotto:2017euk} to a configuration suggesting a natural generalization of the AGT correspondence and incorporating $Y_{N_1,N_2,N_3}$. Such more general truncations correspond to gauge theories supported on the coordinate planes $\mathbb{C}^2_{12},\mathbb{C}^2_{13},\mathbb{C}^2_{23}$ inside $\mathbb{C}^3$ with integers $N_i$ labeling ranks of $U(N_i)$ gauge groups of the corresponding three theories. This generalized version of the AGT correspondence was proven in \cite{Rapcak:2018nsl} (see also \cite{Chuang:2019qdz,Koroteev:2019byp} for related discussion) by showing that algebras $Y_{N_1,N_2,N_3}$ emerge from an action of the cohomological Hall algebra \cite{Kontsevich:2010px} on the equivariant cohomology of the moduli space of spiked instantons from \cite{Nekrasov:2016qym,Nekrasov:2016gud} analogously to the standard $\mathcal{W}_{N}\times \widehat{\mathfrak{gl}(1)}$ story \cite{Schiffmann:2012gf,Maulik:2012wi,Braverman:2014xca}. 

The identification of $Y_{N_1,N_2,N_3}$ with the algebra arising from the action of the cohomological Hall algebra proceeds in two steps. First, one needs to identify the VOA associated to the simplest configurations $Y_{1,0,0}$, $Y_{0,1,0}$ and $Y_{0,0,1}$ corresponding to rank-one theories supported on one of the three coordinate planes. Secondly, one can use an existence of the coproduct structure on $\mathcal{W}_{1+\infty}$ to fuse\footnote{The gauge theories can be physically thought of as arising from the low-energy limit of $N_i$ M5-branes supported on the three coordinate planes. Fusion can be intuitively understood as a process of bringing a system of separated M5-branes together. This motivates the existence of the fusion also in the more complicated setups discussed bellow.} such elementary factors into the general algebra $Y_{N_1,N_2,N_3}$. 

More concretely, each of the elementary factors $Y_{1,0,0}$, $Y_{0,1,0}$ and $Y_{0,0,1}$ can be simply identified with the $\widehat{\mathfrak{gl}(1)}$ current algebra. The coproduct allows us to identify $Y_{N_1,N_2,N_3}$ as a subalgebra of $N_1$ copies of $Y_{1,0,0}$, $N_2$ copies of $Y_{0,1,0}$ and $N_3$ copies of $Y_{0,0,1}$. Such an embedding can be conveniently carved out by a generalized version of the Miura transformation \cite{Fateev:1987zh,Drinfeld:1984qv} for $\mathcal{W}_{N}\times \widehat{\mathfrak{gl}(1)}$.  To define the Miura transformation, \cite{Prochazka:2018tlo} first had to introduce $N_1$ copies of a pseudo-differential operator $\mathcal{L}^{(1)}$, $N_2$ copies of a pseudo-differential operator $\mathcal{L}^{(2)}$ and $N_3$ copies of a pseudo-differential operator $\mathcal{L}^{(3)}$.  By Miura transformation, we mean the process of multiplying these pseudo-differential operators and rewriting them in terms of a pseudo-differential operator in the standard form by commuting derivatives to the right. Generators $U_i$ of the resulting algebra can be then identified with coefficients of such a pseudo-differential operator.

It is natural to ask how general is the relation between VOAs and gauge theories supported on more general complex surfaces \cite{Maulik:2012wi,Negut:2017hxr,Dedushenko:2017tdw,Feigin:2018bkf} or collections of such surfaces wrapping various four-cycles inside higher-dimensional varieties \cite{Rapcak:2019abg,Prochazka:2017qum,Aganagic:2005wn,Jafferis:2006ny,Aganagic:2012si,Nekrasov:2016qym}. It has been expected for a long time that the above-mentioned AGT correspondence only scratches the surface of such a more general story referred to as AGT, 4d/2d or BPS/CFT correspondence by different people. By considering more general configurations, one expects to uncover a rich connection between VOAs and algebraic geometry of the corresponding configuration. 

In this work, we illustrate the richness of the correspondence by analyzing a class of  VOAs corresponding to gauge theories supported on a special class of divisors inside the Calabi-Yau singularity $CY^{3}_{m,n}$ (or its resolutions) given by solutions to the equation
\begin{eqnarray}
xy=z^mw^n
\label{rr}
\end{eqnarray}
for $m,n$ any non-negative integers. The relevant divisors are those that come from zeros of the holomorphic functions of the form
\begin{eqnarray}
x^{N_3} y^{N_2} z^{N_4} w^{N_1}.
\end{eqnarray}

Analogously to the above $Y_{N_1,N_2,N_3}$ algebras, we expect an existence of a definition of the corresponding $x^{N_3} y^{N_2} z^{N_4} w^{N_1}$-algebra in terms of
\begin{enumerate}
\item Truncations of the algebra $\mathcal{W}_{m|n\times\infty}$ generated by an $m|n$ super-matrix of generators for each integral spin $1,2,3,\dots$. 
\item Generalized Miura transformation\footnote{See \cite{tomas} for a detailed discussion of the Miura transformation for $\mathcal{W}_{m|0\times\infty}$ and an identification of some of the truncations by an analysis of null states at low levels.} in terms of a subalgebra of a tensor product of $N_3$ copies of the $x$-algebra, $N_2$ copies of the $y$-algebra, $N_1$ copies of the $z$-algebra and $N_4$-copies of the $w$-algebra. The elementary algebras associated to $x,y,z,w$ will be (roughly) identified with $\widehat{\mathfrak{gl}(m|n)}$ Kac-Moody algebras at specialized levels.
\end{enumerate}
We expect these two definitions to allow an identification of the $x^{N_3} y^{N_2} z^{N_4} w^{N_1}$ VOA with an algebra coming from the geometric action of the cohomological Hall algebra on the equivariant cohomology of the corresponding moduli space of instantons along the lines of \cite{Rapcak:2018nsl}. In particular, $\mathcal{W}_{m|n\times\infty}$ should emerge from doubling the cohomological Hall algebra\footnote{$\mathcal{W}_{m|n\times\infty}$ should emerge as a double of the relevant cohomological Hall algebra. Note that we could consider more general divisors associated to sections of non-trivial bundles on a resolution of $CY^{3}_{m,n}$. We expect these to correspond to truncations of different doubles of the cohomological Hall algebra. Investigation of such more general configurations is left for future work since we do not know how to generalize the Miura transformation for these cases at the moment.} associated to the given geometry $CY^{3}_{m,n}$ and the generalized Miura transformation from its coproduct structure. Understanding the above-mentioned definitions can be thought of as a first step in proving the AGT correspondence for divisors inside $CY^{3}_{m,n}$ that was the main motivation for the project.

Similarly to $\mathcal{W}_{1+\infty}$, algebras $\mathcal{W}_{m|n\times\infty}$ depend on two complex parameters $\Psi$ and $N$ \cite{Prochazka:2017qum,Creutzig:2018pts}. As discussed in the main text, the two parameters can be given a natural geometric interpretation. In particular, the singularity $CY^3_{m,n}$ admits a $T^2$ action preserving the Calabi-Yau volume form. Parameter $\Psi$ then corresponds to a ratio of equivariant parameters associated to such an action. On the other hand, a specialization of the parameter $N$ can be identified with the charge of the function $x^{N_3} y^{N_2} z^{N_4} w^{N_1}$ under such a $T^2$ action.

Finally, note the relation (\ref{rr}) in the ring of holomorphic functions.  We expect the algebras associated to functions $x^{N_3-1} y^{N_2-1} z^{N_4} w^{N_1}$ and $x^{N_3} y^{N_2} z^{N_4+m} w^{N_1+n}$ related by such a relation to be equivalent since they both correspond to the same geometric data. On the other hand, both expressions $x^{N_3} y^{N_2} z^{N_4} w^{N_1}$ and $x^{N_3} y^{N_2} z^{N_4+m} w^{N_1+n}$ lead to a different realization of the algebra in terms of a different subalgebra of a different system of elementary factors. Each relation in the ring of holomorphic functions thus leads to a non-trivial relation between various free-field\footnote{Note that Kac-Moody algebras are not free algebras and embedding inside a tensor product of Kac-Moody algebras is literally not a free-field realization. On the other hand, Kac-Moody algebras can be themselves expressed as subalgebras of products of free bosons, $b,c$ and $\beta,\gamma$ systems that motivates the name ``free-field realization''.} realizations. For example, the simplest relation $xy=z$ for the simplest geometric configuration $CY_{1,0}^3=\mathbb{C}^3$ leads to the well-known bosonization relation that relates the free boson with the $\beta,\gamma$ system. 

The structure of the paper is as follows: Section \ref{geometry} discusses the geometry of $CY^3_{m,n}$. Specializations $m=0$ are known under the name $\mathbb{Z}_n$ singularities and $CY^3_{1,1}$ is known as the conifold singularity. These two cases are extensively discussed at many places in the literature but I have not found a discussion of the more general $CY^3_{m,n}$ case. We describe the toric action mentioned above and the ring of holomorphic functions. Section \ref{webs} discusses resolutions of $CY^3_{m,n}$ and establishes a precise connection to the $(p,q)$-webs colored by integers from \cite{Gaiotto:2017euk,Prochazka:2017qum}. Section \ref{rew} reviews some aspects of the $\mathcal{W}_{1+\infty}$ algebra and the free-field realization of its truncations associated to holomorphic functions on $CY^3_{0,1}=CY^3_{1,0}$. The algebra $\mathcal{W}_{m|n\times \infty}$ generalizing  $\mathcal{W}_{1+\infty}$ is discussed in section \ref{general}. We derive OPEs of weight-one and weight-two fields extending the analysis of \cite{tomas} for the $\mathcal{W}_{m|0\times \infty}$ case. Section \ref{coproduct} introduces the coproduct structure on $\mathcal{W}_{m|n\times \infty}$ generalizing the $\mathcal{W}_{1+\infty}$ case. In section, \ref{blocks} discusses the elementary building blocks of $\mathcal{W}_{m|n\times \infty}$ truncations associated to generators of the ring of holomorphic functions on $CY^3_{m,n}$, i.e. $x,y,z,w$. Finally, section \ref{composition} discusses how to obtain a free-field realization of a general $x^{N_3} y^{N_2} z^{N_4} w^{N_1}$-algebra associated to the function $x^{N_3} y^{N_2} z^{N_4} w^{N_1}$ by composing Miura operators $\mathcal{L}^{(x)},\mathcal{L}^{(y)},\mathcal{L}^{(z)},\mathcal{L}^{(w)}$ associated to the elementary functions. We argue that relations in the ring of holomorphic functions lead to bosonization-like relations between different free-field realizations.

\section{Geometry preliminaries}

\subsection{Calabi-Yau singularities}
\label{geometry}

The main aim of this work is a discussion of truncations of the algebra $\mathcal{W}_{m|n\times \infty}$ that is strongly generated by an $m|n$ super-matrix of generators at each spin $s=1,2,\dots$. Such truncations were conjecturally associated  to a particular class of divisors inside a particular toric Calabi-Yau three-fold (see \cite{Prochazka:2017qum}). Here, we propose that the relevant geometry can be identified with a resolution of the following singularity
\begin{eqnarray}
CY^3_{m,n}=\left \{ (x,y,z,w)\in \mathbb{C}^4|xy=z^mw^n\right \}.
\label{CY}
\end{eqnarray}
This singularity generalizes the well-studied $\mathbb{Z}_m$ singularity and the conifold singularity. In particular, note that for $m=1,n=0$ or $m=0,n=1$, the constraint in (\ref{CY}) becomes simply $z=xy$ or $w=xy$, one can solve for $z$ or $w$ and the corresponding three-fold is just $CY^{3}_{1,0}=CY^{3}_{0,1}=\mathbb{C}^3$ parametrized by the remaining triple of coordinates. More generally, if $n=0$, the coordinate $w$ is unconstrained and $\{(x,y,z)\in \mathbb{C}^3|xy=z^m\}$ is the well-studied $\mathbb{Z}_m$ singularity. We can thus identify $CY^3_{m,0}=\mathbb{C}^2/\mathbb{Z}_m \times \mathbb{C}$. Finally, the case $CY^3_{1,1}$ is known as the conifold singularity. 

The three-fold $CY^3_{m,n}$ admits a natural $T^3$ action that we parametrize as
\begin{eqnarray}
(x,y,z,w)\rightarrow (e^{i(nh_1+(m-1)h_2+mh_3)}x,e^{ih_2}y,e^{i(h_3+h_2)}z,e^{ih_1}w).
\label{action}
\end{eqnarray}
One could in principle choose any other parametrization of the torus that is consistent with the relation in (\ref{CY}). We will see that this choice has a nice interpretation in the $(p,q)$-web picture discussed bellow and $h_i$ correspond to scaling of coordinates in a particular chart on the resolved three-fold. The Calabi-Yau volume form $\Omega$ of $CY^3_{m,n}$ can be expressed in the $x\neq 0$ and the $y\neq 0$ patch as 
\begin{eqnarray}
\Omega_{x\neq 0}=\frac{dy\wedge dz\wedge dw}{y},\qquad \Omega_{y\neq 0}=\frac{dx\wedge dz\wedge dw}{x}.
\end{eqnarray}
The subtorus $T^2 \subset T^3$ preserving the Calabi-Yau volume form then corresponds to the above parameters $h_i$ specialized to
\begin{eqnarray}
h_1+h_2+h_3=0.
\label{hsum}
\end{eqnarray}

\begin{figure}
    \centering
        \includegraphics[width=125mm]{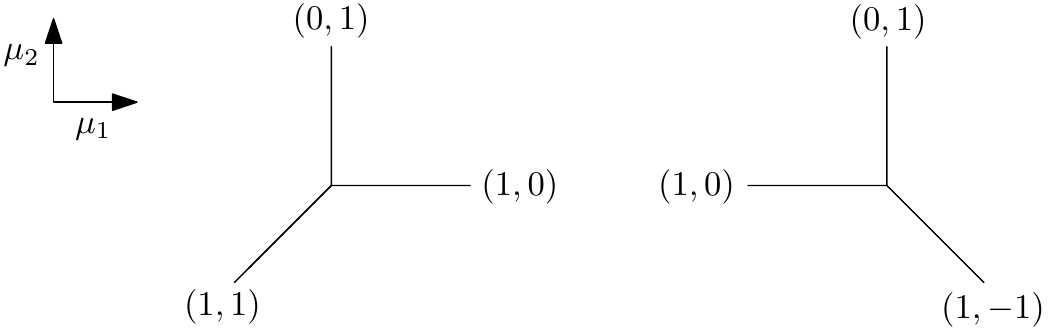}
        \caption{Toric diagram of $CY_{0,1}^3=\mathbb{C}^3$ on the left and $CY_{1,0}^3=\mathbb{C}^3$ on the right.}
        \label{fig1}
\end{figure}

According to the conjecture of \cite{Prochazka:2017qum}, one can associate a vertex operator algebra to any toric divisor inside a toric Calabi-Yau three-fold. We will restrict to a particular class of divisors that can be identified with zero sections of holomorphic functions on $CY^3_{m,n}$. As discussed in section \ref{webs}, resolutions of $CY^3_{m,n}$ lead to configurations from \cite{Prochazka:2017qum}  that are expected to lead to truncations of shifted versions of algebras $\mathcal{W}_{m|n \times \infty}$. The special class of divisors at hand will be identified with configurations with zero shifts and leading to truncations of $\mathcal{W}_{m|n \times \infty}$ itself. The relevant divisors are thus in one-to-one correspondence with functions
\begin{eqnarray}
x^{N_3} y^{N_2} z^{N_4} w^{N_1},
\label{functions}
\end{eqnarray}
for $N_1, N_2,N_3,N_4$ non-negative integers, modulo the relation $xy=z^mw^n$ in the coordinate ring of $CY^3_{m,n}$. Such functions scale with the power of
\begin{eqnarray}
((n-m)h_1-h_2)N_3+h_2N_2 -h_1N_4+h_1N_1
\label{charge}
\end{eqnarray}
under the $T^2$ action described above. This constant will play an important role in the discussion of vertex operator algebras below. Furthermore, note that it reduces to a simple expression
\begin{eqnarray}
h_3 N_3 +h_2 N_2 +h_1N_1
\label{Cspec}
\end{eqnarray}
for $CY^3_{1,0}=\mathbb{C}^3$, where we can use the relation (\ref{CY}) to set $N_4=0$. This specialization is the reason for the above convention for the labels $N_i$ in the exponents of (\ref{functions}).

\subsection{Relation to $(p,q)$-webs}
\label{webs}

\begin{figure}
    \centering
        \includegraphics[width=145mm]{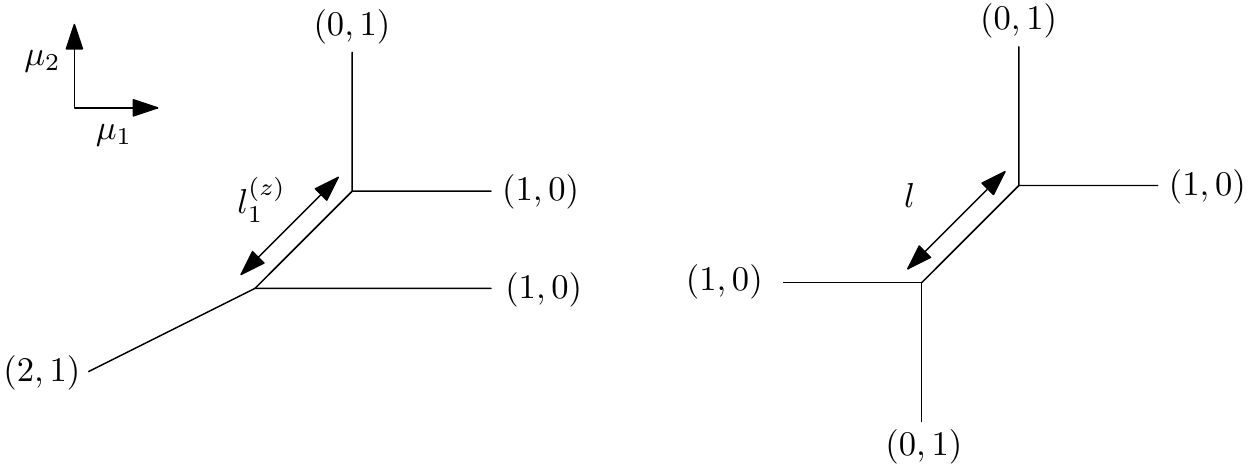}
        \caption{Toric diagram of $\widetilde{CY^3}_{2,0}=\widetilde{\mathbb{C}^2/\mathbb{Z}_2}\times \mathbb{C}$ on the left and the resolved conifold $\widetilde{CY^3}_{1,1}$ on the right.}
        \label{fig2}
\end{figure}

We will now introduce resolutions of $CY^{3}_{m,n}$ and their toric diagram that will allow us to make a contact with the $(p,q)$-webs colored by integers from \cite{Prochazka:2017qum}. For each $CY^{3}_{m,n}$, let us introduce $m+1$ complex coordinates $z_i \in \mathbb{C}$ and $n+1$ complex coordinates $w_i \in \mathbb{C}$. Different triples of $z_i, w_i$ will correspond to coordinates on different patches of the resolved variety. In terms of $z_i,w_i$, one can solve the constraint in (\ref{CY}) and write 
\begin{eqnarray}\nonumber
x&=&z_1^{m}z_2^{m-1}\dots z_{m}w_1^{n}w_2^{n-1}\dots w_{n},\\ \nonumber
y&=&z_{2}z_3^{2}\dots z_{m+1}^mw_{2}w_3^{2}\dots w_{n+1}^n ,\\ \nonumber
z&=&z_1\dots z_n,\\
w&=&w_1\dots w_m.
\label{parametrization}
\end{eqnarray}
There is obviously a lot of freedom in parametrizing $x,y,z,w$ in terms of $z_i,w_i$. In particular, $x,y,z,w$ are invariant with respect to the scaling of each triple $z_i,z_{i+1},z_{i+2}$ or $w_i,w_{i+1},w_{i+2}$ with charges $+1,-2,+1$ and $w_1,w_2,z_{n},z_{n+1}$ with charges $-1,+1,+1,-1$. The corresponding moment maps for such $\mathbb{C}^\times$ actions are\footnote{Note that we have one moment map condition associated to each root of the $\mathfrak{gl}(n)$ and $\mathfrak{gl}(m)$ subalgebra of $\mathfrak{gl}(m|n)$ and one associated to the fermionic root. One can define a different system of moment maps corresponding to a different choice of the root system in $\mathfrak{gl}(m|n)$ and leading to a different resolution of $CY^3_{m,n}$ than the one discussed below.}
\begin{eqnarray}\nonumber
\mu_i^{(z)}&=&|z_i|^2-2|z_{i+1}|^2+|z_{i+2}|^2\quad \mbox{for}\quad i=1,\dots m-1,\\ \nonumber
\mu&=&-|w_1|^2+|w_2|^2+|z_{n}|^2-|z_{n+1}|^2,\\
\mu_j^{(w)}&=&|w_j|^2-2|w_{j+1}|^2+|w_{j+2}|^2\quad \mbox{for}\quad j=1,\dots n-1.
\label{mom}
\end{eqnarray}
The Calabi-Yau singularity $CY^3_{m,n}$ can be then alternatively defined as a symplectic quotinent 
\begin{eqnarray}
CY^3_{m,n}=\mathbb{C}^{m+n+2}//\{\mu^{(z)}_i=\mu=\mu^{(w)}_j=0 \},
\end{eqnarray}
i.e. the preimage of $\mu_i=\mu=\mu_j=0$ inside $\mathbb{C}^{m+n+2}$ modulo $U(1)$ rotations of the coordinates with charges described above.

Calabi-Yau singularity $CY^3_{m,n}$ can be resolved by introducing a real parameter for each moment map $l^{(z)}_i,l,l^{(w)}_j>0$ and considering quotients
\begin{eqnarray}
\widetilde{CY^3}_{m,n}=\mathbb{C}^{m+n+2}//\{\mu_i^{(z)}=l^{(z)}_i,\mu=l,\mu_j^{(w)}=l_j^{(w)} \}.
\end{eqnarray}
Resolved three-folds $\widetilde{CY^3}_{m,n}$ can be thought of as $T^2\times \mathbb{R}$ fibrations over $\mathbb{R}^3$ with the $T^2$ fiber corresponding to the above $h_1,h_2,h_3$ satisfying (\ref{hsum}) and the base parametrized by corresponding moment maps. The geometry of $\widetilde{CY^3}_{m,n}$ can be then encoded in a $(p,q)$-web diagram, i.e. a diagram indicating loci in the base where one of the $T^2$ cycles degenerate \cite{Aganagic:2003db}. The $(p,q)$-web is composed of trivalent vertices associated to different coordinate patches and mutual orientation of vertices then indicate how are two patches glued together.

\begin{figure}
    \centering
        \includegraphics[width=90mm]{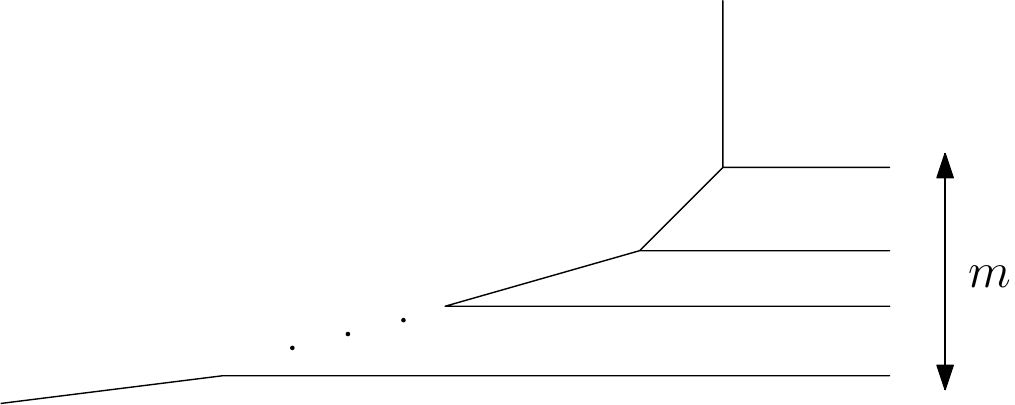}
        \caption{Toric diagram of $\widetilde{CY^3}_{m,0}=\widetilde{\mathbb{C}^2/\mathbb{Z}_2}\times \mathbb{C}$ and half of the toric diagram of the general $\widetilde{CY^3}_{m,n}$.}
        \label{fig3}
\end{figure}

Let us start with a discussion of the toric diagram for $CY_{1,0}^3=\mathbb{C}^3$. The variety can be simply parametrized by $x=z_1,y=z_2,w=w_1$ that scale as
\begin{eqnarray}
(w_1,z_2,z_1)\rightarrow (e^{ih_1}w_1,e^{ih_2}z_2,e^{-i(h_1+h_2)}z_1)
\label{T2C3}
\end{eqnarray}
with the corresponding moment maps
\begin{eqnarray}
\mu_1=|w_1|^2-|z_1|^2,\qquad \mu_2=|z_2|^2-|z_1|^2.
\label{ref1}
\end{eqnarray}
Note that the scaling by $h_1,h_2,h_3$ corresponds to rotations of the three coordinates in our parametrization. This justifies the choice in (\ref{action}).
The diagram indicating degeneration loci of various cycles is depicted in the figure \ref{fig1} on the left. The $(1,0)$ cycle associated to $h_1=0$ degenerates for $z_2=z_1=0$, i.e. along $\mu_2=0,\mu_1\geq 0$, and corresponds to the line $(1,0)$ in the toric diagram. Similarly for the $(0,1)$ cycle associated to $h_2=0$, one gets the $(1,0)$ line. Finally, for the $(1,1)$ cycle associated to $h_1+h_2=0$, one gets the $(1,1)$ line. Considering $CY_{0,1}^3=\mathbb{C}^3$ goes in a similar fashion and produces a diagram with an opposite orientation from the figure \ref{fig1} on the right.

Toric diagram of a more complicated $\widetilde{CY^3}_{m,n}$ can be obtained by gluing more vertices to the simplest trivalent vertex. Let us first illustrate the derivation of the toric diagram for $\widetilde{CY^3}_{2,0}=\widetilde{CY^3}_{0,2}=\widetilde{\mathbb{C}^2/\mathbb{Z}_2}\times \mathbb{C}$ and the resolved conifold $\widetilde{CY^3}_{1,1}$ in detail. After dealing with these two examples, we can directly draw the corresponding diagram for $\widetilde{CY^3}_{m,n}$ since such a general case is a simple combination of the two examples and corresponds to an iterative use of the moment maps (\ref{mom}). 

\begin{figure}
    \centering
        \includegraphics[width=140mm]{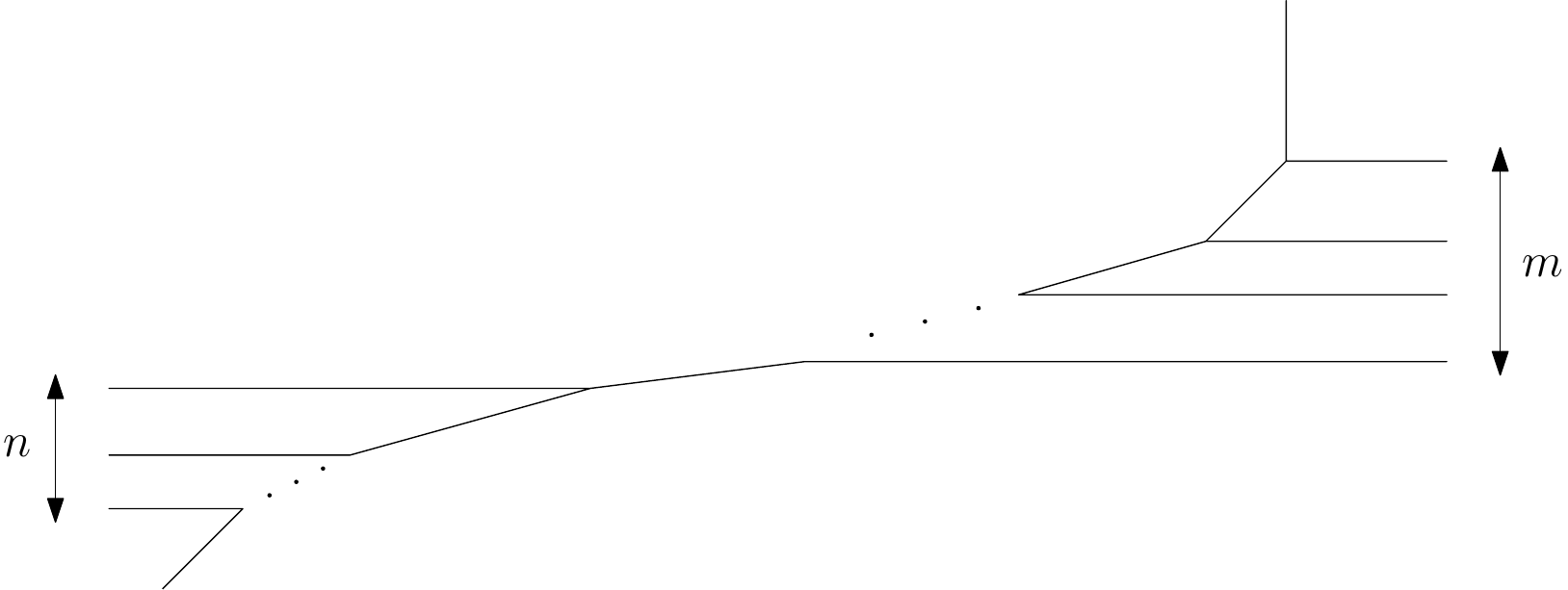}
        \caption{Toric diagram of $\widetilde{CY^3}_{m,n}$.}
        \label{fig4}
\end{figure}

First, in the $\widetilde{CY^3}_{2,0}$ case, we have
\begin{eqnarray}
x=z_1^2z_2,\quad y=z_3^2z_2,\quad z=z_1z_2z_3,\quad w=w_1
\end{eqnarray}
with the corresponding moment map constraint
\begin{eqnarray}
\mu^{(z)}_1=|z_1|^2-2|z_2|^2+|z_3|^2=l^{(z)}_1.
\label{ref3}
\end{eqnarray}
Let us first look at the patch where $z_3\neq 0$. We can then identify the triple $w_1,z_2,z_1$ with coordinates on the corresponding patch with the $T^2$ action as above (\ref{T2C3}). Using the moment map constraint, one can solve for $|z_3|$ and adjust its phase arbitrarily using the corresponding $S^1$ action. We can see that the scaling by $h_1,h_2,h_3$ again corresponds to rotations of the three coordinates in this patch and the corresponding toric diagram in this patch is again the trivalent junction from figure \ref{fig1}. Let us now move to the patch $z_1\neq 0$. The coordinates in this patch are $(w_1,z_3,z_2)$. The moment maps (\ref{ref1}) can be rewritten using (\ref{ref3}) as 
\begin{eqnarray}
\mu_1=|w_1|^2-2|z_2|^2+|z_3|^2-l^{(z)}_1,\qquad \mu_2=-|z_2|^2+|z_3|^2-l^{(z)}_1
\end{eqnarray}
with the corresponding fiber generated by
\begin{eqnarray}
(w_1,z_3,z_2)\rightarrow (e^{ih_1}w_1,e^{i(h_1+h_2)}z_3,e^{-i(2h_1+h_2)}z_2).
\end{eqnarray}
Note first that the corresponding trivalent junction associated to this patch is shifted by $(-l^{(z)}_1,-l^{(z)}_1)$ in the $(\mu_1,\mu_2)$ plane. Furthermore, we see that in this patch, we get degenerations along the lines $(1,0),(1,1)$ and $(2,1)$ with $h_1=0,h_1+h_2=0,2h_1+h_2=0$ being the degenerating cycles. The corresponding toric diagram of  $\widetilde{CY^3}_{2,0}$ is shown in the figure \ref{fig2} on the left. We get an analogous diagram for $\widetilde{CY^3}_{0,2}$ with an opposite orientation similarly to $\widetilde{CY^3}_{0,1}$.

\begin{figure}
    \centering
        \includegraphics[width=22mm]{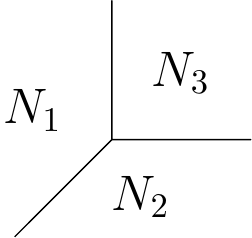}
        \caption{A colored $(p,q)$-web specifying divisor of $x^{N_3}y^{N_2}w^{N_1}$ in $CY^3_{1,0}=\mathbb{C}^3$.}
        \label{fig5}
\end{figure}

For $\widetilde{CY^3}_{1,1}$, we have
\begin{eqnarray}
x=z_1w_1,\quad y=z_2w_2,\quad z=z_1z_2,\quad w=w_1w_2.
\end{eqnarray}
with the corresponding moment map
\begin{eqnarray}
\mu=|z_1|^2-|z_2|^2-|w_1|^2+|w_2|^2=l.
\label{ref2}
\end{eqnarray}
First, the patch $w_2\neq 0$ can be again parametrized by the triple $(w_1,z_2,z_1)$ leading to one trivalent vertex with $h_i$ rotating the three coordinates. In the chart $z_1\neq 0$, we have coordinates $(w_1,z_2,w_2)$ with the corresponding moment maps
\begin{eqnarray}
\mu_1=-|z_2|^2+|w_2|^2-l,\qquad \mu_2=-|w_1|^2+|w_2|^2-l
\end{eqnarray}
and the associated toric action
\begin{eqnarray}
(w_1,z_2,w_2)\rightarrow(e^{-ih_2}w_1,e^{-ih_1}z_2,e^{i(h_1+h_2)}w_2).
\end{eqnarray}
We can see that this leads to a vertex shifted by $(-l,-l)$ in the $(\mu_1,\mu_2)$ plane with $(1,0)$, $(0,1)$, $(1,1)$ cycles degenerating as shown in the figure \ref{fig2} on the right.

Finally, let us describe a general $(p,q)$-web associated to a general $\widetilde{CY^3}_{m,n}$. As in the previous examples, one can again associate one vertex with the $(w_1,z_2,z_1)$ coordinates corresponding to the patch where all the other $z_i$ and $w_i$ are non-vanishing. We can then use the moment map conditions $\mu^{(z)}_i$ to find moment maps on patches parametrized by triples $(w_1,z_{i+1},z_i)$ giving rise to a sequence of vertices from the figure \ref{fig3}. The moment map condition $\mu$ attaches a reversed vertex with the $(1,0)$ line on the left to the last line of figure \ref{fig3} as in the resolved conifold example. Finally, using moment maps $\mu^{(w)}_i$, one can find expressions for $\mu_1$ and $\mu_2$ on patches parametrized by the triple $(w_i,z_{n+1},w_{i+1})$. The resulting $(p,q)$-web is shown in the figure \ref{fig4}.

\begin{figure}
    \centering
    \begin{subfigure}[b]{0.38\textwidth}
        \includegraphics[width=\textwidth]{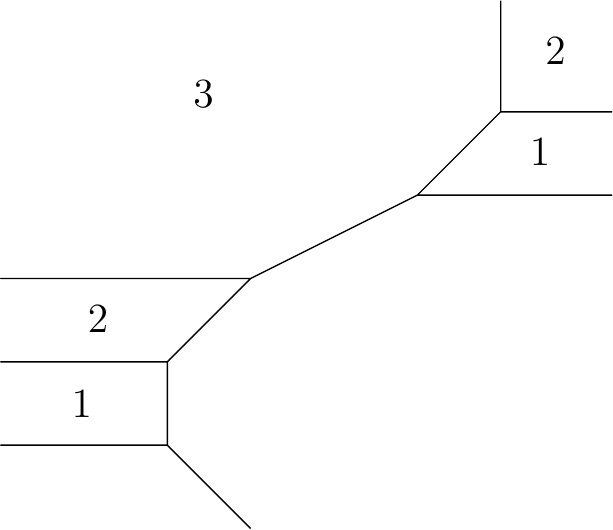}
        \caption{A divisor of $x$.}
        \label{refx}
    \end{subfigure}\qquad
    \begin{subfigure}[b]{0.38\textwidth}
        \includegraphics[width=\textwidth]{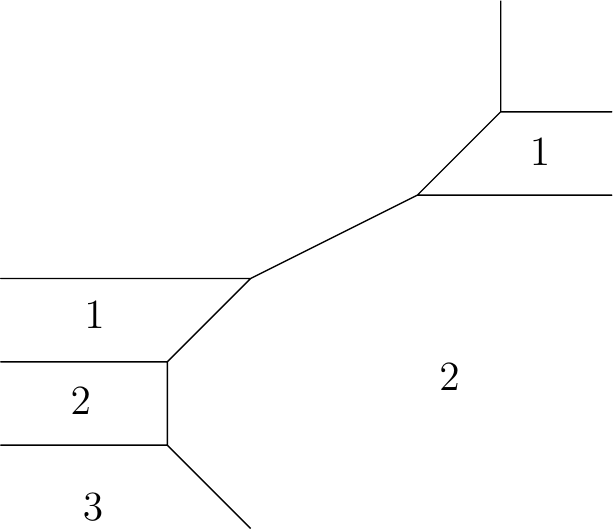}
        \caption{A divisor of $y$.}
        \label{refy}
    \end{subfigure}\\
    \begin{subfigure}[b]{0.38\textwidth}
        \includegraphics[width=\textwidth]{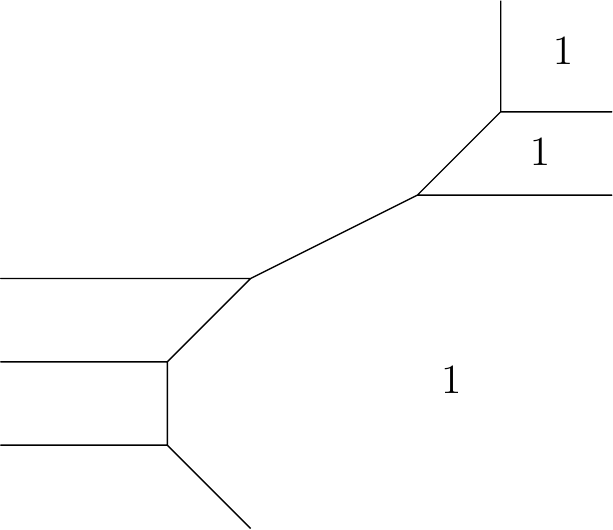}
        \caption{A divisor of $z$.}
        \label{refz}
    \end{subfigure}\qquad
    \begin{subfigure}[b]{0.38\textwidth}
        \includegraphics[width=\textwidth]{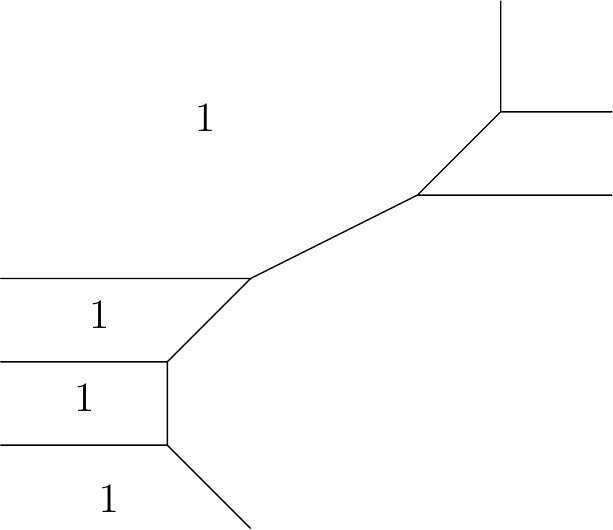}
        \caption{A divisor of $w$.}
        \label{refw}
    \end{subfigure}
    \caption{Divisors associated to the elementary functions $x,y,z,w$ in the resolved three-fold $\widetilde{CY^3}_{2,3}$.}\label{factors}
\end{figure}

Let us now discuss functions (\ref{functions}) and divisors associated to their zero sections from the perspective of the $(p,q)$-web. First, considering the simplest case $CY^3_{1,0}=\mathbb{C}^3$, the coordinate planes $x=0$, $y=0$, $w=0$ map to the three corners of the toric diagram under the maps (\ref{ref1}). The divisor of a general function $x^{N_3}y^{N_2}w^{N_1}$ can be then labeled by the colored $(p,q)$ web from the figure \ref{fig5} with integers $N_1,N_2,N_3$ indicating multiplicities of different smooth components of the divisor.

This discussion has a generalization to the general case of $CY^{3}_{m,n}$. Each function $x,y,w,z$  can be rewritten in terms of $z_i,w_i$ using (\ref{parametrization}). Restricting to a patch associated to a given vertex, we can use the moment map conditions \ref{ref2} to solve for $|z_i|$ and $|w_i|$ that are not part of the corresponding triple and remove the phase of the corresponding $z_i,w_i$. One can then read off the corresponding multiplicities of the smooth components in the given patch from powers of the corresponding $z_i,w_i$ and deduce integers for the corresponding vertex in the $(p,q)$-web.  This leads globally to a colored $(p,q)$-webs with integers associated to each face of the $(p,q)$-web. Divisors associated to the elementary functions $x,y,z,w$ from the $(p,q)$-web perspective are shown for the example of $CY^3_{2,3}$ in figure \ref{factors}. Generally, $x$ leads to a diagram with decreasing numbers from the top to the bottom starting with $m$ on the right and $n$ on the left. On the other hand, $y$ corresponds to decreasing numbers from the bottom to the up again starting with $m$ on the right and $n$ on the left. $z$ is associated with ones on the right-hand side of the diagram and $w$ with ones on the left.

\begin{figure}
    \centering
        \includegraphics[width=100mm]{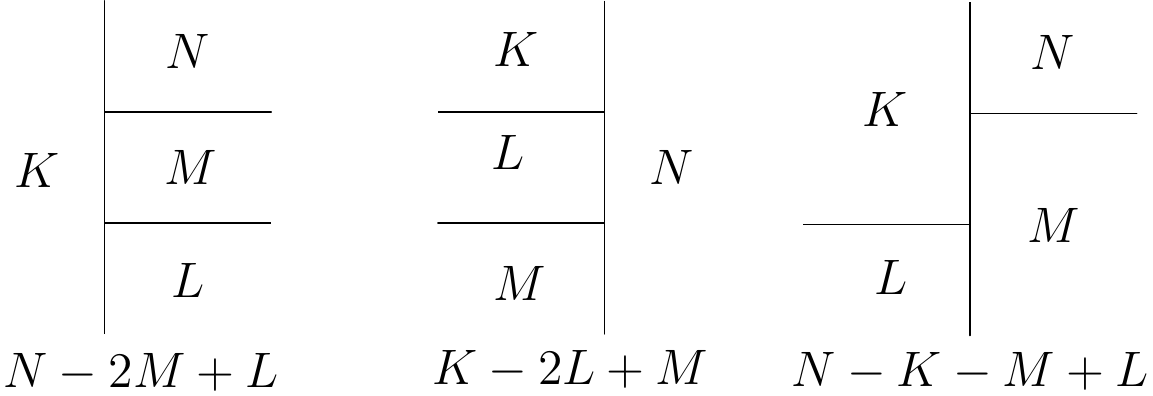}
        \caption{Shift parameters associated to different finite segments of the diagram.}
        \label{fig6}
\end{figure}

Let us conclude this section with two remarks. First, multiplication of holomorphic functions corresponds to summing corresponding divisors and in particular summing multiplicities associated to various faces of the toric diagram. The relation $xy=z^mw^n$ then equates two different ways of producing the same divisor using such sums. On the VOA side, this multiplication will lead to a coproduct encoded in a generalization of the Miura transformation. Secondly, note that \cite{Prochazka:2017qum} conjectured that algebras associated to the web \ref{fig4} should be truncations of shifted versions of $\mathcal{W}_{m|n \times \infty}$ with shifting parameters associated to each finite segment of the $(p,q)$-web. The shift parameters associated to various configurations in our setup are schematically depicted in figure \ref{fig6}. They can be identified with the $U(1)$ charge of the divisor with respect to the action generated by moment maps (\ref{mom}). Note that all the configurations associated to $x^{N_3}y^{N_2}z^{N_4}w^{N_1}$ have zero shift. Furthermore, it is not hard to convince yourself that all zero-shift configurations come from some $x^{N_3}y^{N_2}z^{N_4}w^{N_1}$.

\section{Review of the $\mathbb{C}^3$ case}
\label{rew}

Let us now move to the discussion of vertex operator algebras and review the generalized Miura transformations from \cite{Prochazka:2018tlo}. $\mathcal{W}_N\times \widehat{\mathfrak{gl}(1)}$ algebras admit a well-known realization \cite{Fateev:1987zh,Drinfeld:1984qv,Prochazka:2014aa,Genra:2019tgw} in terms of a subalgebra of $N$ copies of $\widehat{\mathfrak{gl}(1)}$. Let us normalize the corresponding strong generators $J^{(x)}_i$ of $\widehat{\mathfrak{gl}(1)}$ for $i=1,\dots,N$ as
\begin{eqnarray}
J^{(x)}_i(z)J^{(x)}_i(w)\sim \frac{\kappa+1}{(z-w)^2}.
\end{eqnarray}
We can then define Miura operators as formal expressions
\begin{eqnarray}
\mathcal{L}^{(x)}_i=\kappa\partial + J^{(x)}_i.
\end{eqnarray}
Considering a product of such operators and rewriting it as an $N$'th order differential operator in the standard form
\begin{eqnarray}
\mathcal{L}^{(x)}_1\mathcal{L}^{(x)}_2 \dots \mathcal{L}^{(x)}_N = (\kappa\partial)^N+U_1(\kappa\partial)^{N-1}+U_1(\kappa\partial)^{N-2}+\dots +U_N
\end{eqnarray}
by commuting all derivatives to the right, one obtains strong generators $U_i$  of the $\mathcal{W}_N \times \widehat{\mathfrak{gl}(1)}$ algebra in so-called $U$-basis (sometimes also called the quadratic basis since OPEs in this basis are quadratic in derivatives of $U_i$'s). In particular, we have
\begin{eqnarray}\nonumber
U_1&=&\sum_{i=1}^N J^{(x)}_i\\ \nonumber
U_2&=&\sum_{i=1}^N\sum_{j=i+1}^N (J^{(x)}_iJ^{(x)}_j)+\kappa \sum_{i=1}^N (i-1)\partial J^{(x)}_i\\ \nonumber
U_3&=&\sum_{i=1}^N\sum_{j=i+1}^N\sum_{k=j+1}^N (J^{(x)}_iJ^{(x)}_jJ^{(x)}_k)+\kappa\sum_{i=1}^N\sum_{j=i+1}^N (i-1)(\partial J^{(x)}_iJ^{(x)}_j)\\
&&+\kappa\sum_{i=1}^N\sum_{j=i+1}^N (j-2)( J^{(x)}_i\partial J^{(x)}_j)+\frac{\kappa^2}{2}\sum_{i=1}^N (i-1)(i-2)\partial J^{(x)}_i
\end{eqnarray}
that leads to an algebra with first few OPEs of the form
\begin{eqnarray}\nonumber
U_1(z)U_1(w)&\sim& \frac{N(\kappa+1)}{(z-w)^2},\\ \nonumber
U_1(z)U_2(w)&\sim& \frac{N(N-1)\kappa (\kappa+1)}{(z-w)^3}+\frac{(N-1)(\kappa+1)U_1}{(z-w)^2},\\ \nonumber
U_1(z)U_3(w)&\sim& \frac{N(N-1)(N-2)\kappa^2 (\kappa+1)}{(z-w)^3}+\frac{(N-1)(N-2)\kappa (\kappa+1)U_1}{(z-w)^3}\\ \nonumber
&&+\frac{(N-2)(\kappa+1)U_2}{(z-w)^2},\\ \nonumber
U_2(z)U_2(w)&\sim& \frac{N(N-1)(\kappa+1)(1+\kappa+2(2N-1)\kappa^2)}{2(z-w)^4}\\ \nonumber
&&-\frac{2U_2-(N-1)(\kappa+1)U_1U_1-\kappa (\kappa+1) N(N-1)\partial U_1}{(z-w)^2}\\
&&-\frac{\kappa+1}{2}\frac{2\partial U_2-2(N-1)(\partial U_1U_1)-\kappa  N(N-1)\partial^2 U_1}{z-w}.
\label{WinfOPE}
\end{eqnarray}
Considering the large $N$ limit and treating the parameter $N$ as a generic complex number, one obtains an algebra $\mathcal{W}_{1+\infty}$ generated by fields $U_i$ depending on two continuous parameters $\kappa$ and $N$ with the above OPEs. It was argued in \cite{Gaberdiel:2012aa,Prochazka:2014aa,Linshaw:2017tvv} that this algebra is actually uniquely fixed by requiring the associativity of OPEs and an existence of a single strong generator for each integral spin. 

We can then look at the algebras $\mathcal{W}_N\times \widehat{\mathfrak{gl}(1)}$ as truncations of such a two-parameter family of algebras $\mathcal{W}_{1+\infty}$ coming from the specialization $N\in \mathbb{Z}$ to be an integer and setting $U_i=0$ for $i>N$. It was argued in \cite{Prochazka:2017qum} that there actually exists a larger family of truncations $Y_{N_1,N_2,N_3}$ parametrized by three integral numbers $N_1,N_2,N_3$. To describe the specialization of the parameter $N$ associated to such truncations, it is useful to introduce parameters $h_1,h_2,h_3$ such that\footnote{The relation between $\Psi$ and $\kappa$ will be modified for more general $CY^3_{m,n}$.}
\begin{eqnarray}
h_1+h_2+h_3=0,\quad \kappa+1=\Psi=-\frac{h_2}{h_1}.
\end{eqnarray}
We can then write the formula for the specialized $N$ as
\begin{eqnarray}
N=\frac{N_1h_1+N_2h_2+N_3h_3}{h_3}.
\label{Winfspec}
\end{eqnarray}
Note that the numerator equals the charge (\ref{Cspec}) for the transformation of holomorphic functions on $CY^3_{1,0}=\mathbb{C}^3$ with respect to the $T^2$ action. Furthermore, the expression is independent of the overall scale of $h_i$ and the resulting algebra in the end depends only on the ratio $\Psi$. We will see a natural generalization of the formula for $CY^3_{m,n}$ bellow.

It was found in \cite{Prochazka:2018tlo} that there actually exist two more Miura operators whose products give rise to $Y_{N_1,N_2,N_3}$ truncations. Let us normalize the currents appearing in these two Miura operators as\footnote{Note that the labeling $x,y,w$ is in correspondence with generators of the ring of holomorphic functions on $CY^3_{1,0}=\mathbb{C}^3$ from the previous section.}
\begin{eqnarray}\nonumber
J^{(w)}(z)J^{(w)}(w)&\sim&\frac{\kappa+1}{\kappa}\frac{1}{(z-w)^2},\\
J^{(y)}(z)J^{(y)}(w)&\sim&-\frac{(\kappa+1)^2}{\kappa}\frac{1}{(z-w)^2}.
\end{eqnarray}
It is a simple calculation to identify combinations of normally-ordered products of derivatives of $J^{(w)}$ and $J^{(y)}$ respectively that satisfy OPEs (\ref{WinfOPE}) with the parameter $N$ specialized according to (\ref{Winfspec}) as $(N_1,N_2,N_3)=(1,0,0)$ in the first case and as $(N_1,N_2,N_3)=(0,1,0)$ in the second case. One gets
\begin{eqnarray}\nonumber
U^{(w)}_1&=&J^{(w)},\\
U^{(w)}_2&=&\frac{1-\kappa}{2}\left ((J^{(w)}J^{(w)})+\partial J^{(w)}\right ),\\  \nonumber
U^{(w)}_3&=&\frac{1}{6}\left (\kappa-1\right )\left (2\kappa-1\right )\left ((J^{(w)}J^{(w)}J^{(w)})+3(J^{(w)}\partial J^{(w)})+\partial^2 J^{(w)}\right )
\end{eqnarray}
and
\begin{eqnarray}\nonumber
U^{(y)}_1&=&J^{(y)},\\ 
U^{(y)}_2&=&\frac{2\kappa+1}{2(\kappa+1)}\left ((J^{(y)}J^{(y)})-(\kappa+1)\partial J^{(y)}\right ),\\ \nonumber
U^{(y)}_3&=&\frac{(2\kappa+1)(3\kappa+1)}{6(\kappa+1)^2}\left ((J^{(y)}J^{(y)}J^{(y)})-3(\kappa+1)(J^{(y)}\partial J^{(y)})+(\kappa+1)^2\partial^2 J^{(y)}\right ).
\end{eqnarray}

Having identified $U$-generators of the three elementary truncations, we can then define a generalized Miura operator as\footnote{A simple and uniform expression for $\mathcal{L}^{(x)},\mathcal{L}^{(y)},\mathcal{L}^{(w)}$ in terms of a conjugation of $\partial^{\frac{h_i}{h_3}}$ by an exponential operator appeared in \cite{Prochazka:2019dvu}. We do not use this representation here since it is not obvious how to generalize such expressions to the matrix case described below. It would be nice to check whether such a generalization exists.}
\begin{eqnarray}
\mathcal{L}^{(\alpha)}=(\kappa\partial)^{N_{\alpha}}+U^{(\alpha)}_{1}(\kappa\partial)^{N_\alpha-1}+U^{(\alpha)}_{2}(\kappa\partial)^{N_\alpha-2}+\dots,
\end{eqnarray}
where $\alpha=x,y,w$ and $N_\alpha$ is the correctly-specialized parameter $N$, i.e. $N_x=1$ for the $x$-truncation, $N_y=\frac{h_2}{h_3}$ for the $y$-truncation and $N_w=\frac{h_1}{h_3}$ for the $w$-truncation. Multiplying $N_1$ elementary factors $\mathcal{L}^{(w)}_i$, $N_2$ elementary factors $\mathcal{L}^{(y)}_i$ and $N_3$ elementary factors $\mathcal{L}^{(x)}_i$ and commuting derivatives to the right using the Leibniz rule, we recover a pseudo-differential operator
\begin{eqnarray}
\partial^{N}(fg)= f\partial^{N}g+N\partial f \partial^{N-1}g+N(N-1)\partial^2 f \partial^{N-2}g+\dots,
\label{com_der}
\end{eqnarray}
whose coefficients $U_i$ give the free-field realization of generators of the algebra. Note that $N$ is additive under such a multiplication and its value for a general truncation agrees with (\ref{Winfspec}). 

After setting the notation for $\mathcal{W}_{1+\infty}$ that is convenient to make contact with the geometric picture from the previous section, we will move to a matrix generalization of the story. As we will see, the interplay between the geometry and properties of VOAs become much more non-trivial in the matrix case.

\section{$\mathcal{W}_{m|n \times \infty}$ algebra}
\label{general}

We will now discuss a matrix generalization of the $\mathcal{W}_{1+\infty}$ algebra, i.e. an algebra $\mathcal{W}_{m|n\times \infty}$ generated by $m|n$ super-matrices of fields. Analogously to the $\mathcal{W}_{1+\infty}$ case, we expect that there exists a two-parameter family of algebras $\mathcal{W}_{m|n \times \infty}$ satisfying the following conditions:
\begin{enumerate}
\item $\mathcal{W}_{m|n \times \infty}$ contains the Kac-Moody algebra $\widehat{\mathfrak{gl}(m|n)}_{\tilde{\kappa}}$ at level $\tilde{\kappa}$ as a subalgebra.
\item $\mathcal{W}_{m|n \times \infty}$ is strongly generated by a $m|n$ super-matrix of generators $U_{i,ab}$ at each integral spin $i=1,2,3,\dots$. Indices $a,b$ label the row and the column index of the super-matrix with $a\leq m$ bosonic and $a>m$ fermionic directions, such that the $n\times n$ and the $m\times m$ diagonal block consists of bosonic generators with the off-diagonal blocks being fermionic. 
\item The generators $U_{i,ab}$ at given spin $i$ transform in the adjoint representation of the $\mathfrak{gl}(m|n)$ zero-mode subalgebra of $\widehat{\mathfrak{gl}(m|n)}_{\tilde{\kappa}}$.
\end{enumerate}
In analogy with the $\mathcal{W}_{1+\infty}$ algebra, it is tempting to speculate that $\mathcal{W}_{m|n \times \infty}$ is uniquely fixed by Jacobi identities up to the two parameters.

Authors of \cite{tomas} argued that one can easily find OPEs of $\mathcal{W}_{m|0\times \infty}$ in a generalized $U$-basis using a matrix generalization of the Miura transformation. In this section, we will use the same method to determine OPEs of low-spin generators of its super-matrix generalization $\mathcal{W}_{m|n\times \infty}$. 

Our starting point is the $\widehat{\mathfrak{gl}(m|n)}_\kappa$ Kac-Moody algebra generated by fields $J^{(x)}_{ab}$ with OPE
\begin{eqnarray}
J_{ab}^{(x)}(z)J_{cd}^{(x)}(w) &\sim&  \frac{(-1)^{|b||c|}\kappa \delta_{ad}\delta_{cb}+\delta_{ab}\delta_{cd}}{(z-w)^2}\\ \nonumber
&&+\frac{(-1)^{|a||b|+|c||d|+|c||b|}\delta_{ad}J_{cb}^{(x)}-(-1)^{|b||c|}\delta_{cb}J_{ad}^{(x)}}{z-w}.
\end{eqnarray}
$\widehat{\mathfrak{gl}(m|n)}_\kappa$ is clearly a truncation of $\mathcal{W}_{m|n \times \infty}$ in the same way as $\widehat{\mathfrak{gl}(1)}$ is the simplest truncation of $\mathcal{W}_{1+\infty}$. In the above OPE, we have introduced a continuous parameter 
\begin{eqnarray}
\kappa=\Psi-m+n
\end{eqnarray}
and $|a|=0$ for fermionic directions $a\leq m$ and one otherwise. For the notational convenience, we often omit the brackets $|a|$ in exponents from now on.

One can then define the Miura operator
\begin{eqnarray}
\mathcal{L}^{(x)}_i=\kappa\partial +J^{(x)}_i,
\end{eqnarray}
where $J^{(x)}_i$ is an $m|n$ super-matrix with $J^{(x)}_{i,ab}$ entries. We can then consider a product of $N$ such factors and rewrite them as a super-matrix-valued differential operator in the standard form
\begin{eqnarray}
\mathcal{L}^{(x)}_1\mathcal{L}^{(x)}_2 \dots \mathcal{L}^{(x)}_N =(\kappa\partial)^N+U_1(\kappa\partial)^{N-1}+U_2(\kappa\partial)^{N-2}+\dots
\end{eqnarray}
with $J^{(x)}_i$ multiplying as super-matrices and $U_i$ being super-matrices as well. In particular, we have in terms of components
\begin{eqnarray}\nonumber
U_{1,ab}&=& \sum_{i=1}^{N}J^{(x)}_{i,ab},\\ \nonumber
U_{2,ab}&=& \sum_{i=1}^{N} \sum_{j=i+1}^{N}\sum_{c}(J^{(x)}_{i,ac}J^{(x)}_{j,cb})+\kappa  \sum_{i=1}^{N}(i-1)\partial J^{(x)}_{i,ab}, \\ \nonumber
U_{3,ab}&=& \sum_{i=1}^{N} \sum_{j=i+1}^{N}\sum_{k=j+1}^{N}\sum_{cd}(J^{(x)}_{i,ac}(J^{(x)}_{j,cd}J^{(x)}_{k,db}))+\kappa \sum_{i=1}^N\sum_{j=i+1}^N(i-1)\sum_c(\partial J^{(x)}_{i,ac}J^{(x)}_{j,cb} )\\
&&+\kappa \sum_{i=1}^N\sum_{j=i+1}^N(j-2)\sum_c ( J^{(x)}_{i,ac}\partial J^{(x)}_{j,cb} )+\frac{\kappa^2}{2}\sum_{i=1}^N (i-1)(i-2)\partial^2 J_{i,ab}^{(x)}.
\end{eqnarray}

Using these expressions, it is straightforward (though technically demanding) to determine OPEs of these fields. For the first few fields, we get\footnote{Note that the expressions  reduce to (\ref{WinfOPE}) for $m|n=1|0$.}
\begin{eqnarray}\nonumber
U_{1,ab}(z)U_{1,cd}(w)&\sim&\frac{N((-1)^{bc}\kappa \delta_{ad}\delta_{cb}+\delta_{ab}\delta_{cd})}{(z-w)^2}+\frac{(-1)^{ab+cd+cb}\delta_{ad}U_{1,cb}-(-1)^{bc}\delta_{db}U_{1,ad}}{z-w},  \\ \nonumber
U_{1,ab}(z)U_{2,cd}(w)&\sim&\frac{N(N-1)\kappa ((-1)^{bc}\kappa \delta_{ad}\delta_{cb}+\delta_{ab}\delta_{cd})}{(z-w)^3}\\ \nonumber
&&+\frac{(N-1)((-1)^{ab+cd+cb}\kappa \delta_{ad}U_{1,cb}+\delta_{ab}U_{1,cd})}{(z-w)^2}\\ \nonumber
&&+\frac{(-1)^{ab+cd+cb}\delta_{ad}U_{2,cb}-(-1)^{bc}\delta_{cb}U_{2,ad}}{z-w}, \\ \nonumber
U_{1,ab}(z)U_{3,cd}(w)&\sim& \frac{N(N-1)(N-2)\kappa^2 ((-1)^{bc}\kappa \delta_{ad}\delta_{cb}+\delta_{ab}\delta_{cd})}{(z-w)^4}\\ \nonumber
&&+\frac{(N-1)(N-2)\kappa ((-1)^{ab+cd+cb}\kappa \delta_{ad}U_{1,cb}+\delta_{ab}U_{1,cd})}{(z-w)^3}\\ \nonumber
&&+\frac{(N-2) ((-1)^{ab+cd+cb}\kappa \delta_{ad}U_{2,cb}+\delta_{ab}U_{2,cd})}{(z-w)^2}\\ \nonumber
&&+\frac{(-1)^{ab+cd+cb}\delta_{ad}U_{3,cb}-(-1)^{bc}\delta_{cb}U_{3,ad}}{z-w}, \\ \nonumber
U_{2,ab}(z)U_{2,cd}(w)&\sim& \frac{N(N-1)}{2}\frac{2\kappa ((1-2N)\kappa^2 +1)(-1)^{bc}\delta_{ad}\delta_{bc}-((4N-3)\kappa^2-1)\delta_{ab}\delta_{cd}}{(z-w)^4}\\ \nonumber
&&+\frac{(N-1)^2\kappa (\delta_{cd}U_{1,ab}-\delta_{ab}U_{1,cd})}{(z-w)^3}\\ \nonumber
&&+\frac{(N-1)(N\kappa^2-1)((-1)^{bc}\delta_{cb}U_{1,ad}-(-1)^{ab+cd+cb}\delta_{ad}U_{1,cb})}{(z-w)^3}\\ \nonumber
&&-\frac{\delta_{ab}U_{2,cd}+\delta_{cd}U_{2,ab}+\kappa ((-1)^{bc}\delta_{cb}U_{2,ad}+(-1)^{ab+ac+bc}\delta_{ad}U_{2,cb})}{(z-w)^2}\\ \nonumber
&&+\frac{(N-1)((U_{1,ab}U_{1,cd})+(-1)^{bd+cd+bc}\kappa (U_{1,ad}U_{1,cb}))}{(z-w)^2}\\ \nonumber
&&+\frac{N(N-1)\kappa (-1)^{bc}(\delta_{cd}\partial U_{1,ab}+\kappa \delta_{cb}\partial U_{1,ad})}{(z-w)^2}\\ \nonumber
&&-\frac{(-1)^{bc}\delta_{cb}U_{3,ad}-(-1)^{ab+ac+bc}\delta_{ad}U_{3,cb}}{z-w}\\ \nonumber
&&+\frac{\delta_{ab}\partial U_{2,cd}-\delta_{cd}\partial U_{2,ab}-((-1)^{bc}\kappa \delta_{cb}\partial U_{2,ad} +\delta_{cd}\partial U_{2,ab})}{z-w}\\ \nonumber
&&+\frac{(-1)^{ab+ac+bc}(U_{2,cb}U_{1,ad})-(-1)^{bd+cd+bc}(U_{2,ad}U_{1,cb})}{z-w}\\ \nonumber
&&+\frac{(N-1)((\partial U_{1,ab}U_{1,cd})+\kappa (-1)^{bd+cd+bc}(\partial U_{1,ad}U_{1,cb}))}{z-w}\\ \nonumber
&&+\frac{N-1}{2}\frac{(N+1)\kappa \delta_{cd}\partial^2 U_{1,ab}-\kappa \delta_{ab}\partial^2 U_{1,cd}}{z-w}\\
&&+\frac{N-1}{2}\frac{(-1)^{bd+cd+bc}((N \kappa^2+1)\delta_{cb}\partial^2 U_{1,ad}-\delta_{ad}\partial^{2}U_{1,cb})}{z-w}.
\label{allOPES}
\end{eqnarray}
These four OPEs are going to be sufficient for testing our proposals for elementary Miura operators and the generalized Miura transformation below. 

We can now state the following conjecture: 
\begin{conj}
There exists a VOA satisfying conditions listed above and containing fields $U_{1,ab},U_{2,ab},U_{3,ab}$ satisfying (\ref{allOPES}) for any $\Psi,N\in \mathbb{C}$. In analogy with $\mathcal{W}_{1+\infty}$, we conjecture that the algebra is uniquely fixed by associativity of OPEs. 
\end{conj}

Let us finish this section by making few comments on the relation to other work. Structure constants of $\mathcal{W}_{m|n\times \infty}$ are up to the $(-1)^{\#}$ factors the same as those in \cite{tomas}. It should be straightforward to find a proposal for all  OPEs of $\mathcal{W}_{m|n\times \infty}$ by correctly recovering these $(-1)^{\#}$ factors in the formulas of \cite{tomas}. Algebras of type $m|n=m|0$ have been recently also studied in a different basis for example in \cite{Costello:2016nkh,Prochazka:2017qum,Creutzig:2018pts,Creutzig:2019qos} who used names rectangular $\mathcal{W}$-algebras or $\mathcal{W}_{m+\infty}$. Furthermore, the algebra $\mathcal{W}_{1+\infty}$ is well-known to be isomorphic to the affine Yangian of $\mathfrak{gl}(1)$ \cite{Schiffmann:2012gf,Prochazka:2015deb}. We expect our $\mathcal{W}_{m|n \times \infty}$ to be isomorphic to the affine Yangian of the Lie super-algebra $\mathfrak{gl}(m|n)$ discussed from different perspectives for example in \cite{Feigin:2013fga,Bershtein:2019shb,Gaberdiel:2017hcn,Gaberdiel:2018nbs,Li:2019nna}.

\section{Miura transformation for $CY^3_{m,n}$}

So far, we have used the simple Miura operator of the form 
\begin{eqnarray}
\mathcal{L}^{(x)}_i=\kappa \partial+J^{(x)}_i,
\label{sim_miu}
\end{eqnarray}
where $J^{(x)}_i(z)$ is a super-matrix of the $\widehat{\mathfrak{gl}(m|n)}_{\kappa}$ Kac-Moody generators to determine OPEs of the $\mathcal{W}_{m|n\times \infty}$ algebra in the $U$-basis. In this section, we generalize the above discussion in two ways. First, we introduce more general matrix-valued pseudo-differential Miura operators to define coproduct on $\mathcal{W}_{m|n\times \infty}$. The composition of $\mathcal{L}^{(x)}_i$ can be then interpreted as a successive use of such a coproduct for the simplest truncation of the algebra. Secondly, we use the geometric and gauge-theoretical picture to determine four elementary truncations associated to functions $x,y,z,w$. All the other truncations associated to $x^{N_3}y^{N_2}z^{N_4}w^{N_1}$ can be then constructed using the coproduct of $\mathcal{W}_{m|n\times \infty}$ and the knowledge of the elementary factors.

\subsection{Coproduct of $\mathcal{W}_{m|n\times \infty}$}
\label{coproduct}

The Miura operator coming from the product of $N$ elementary Miura factors $\mathcal{L}^{(x)}_i$ has generally the following form
\begin{eqnarray}
\mathcal{L}=(\kappa\partial)^N+U_{1}(\kappa\partial)^{N-1}+U_{2}(\kappa\partial)^{N-2}+\dots +U_{N}
\end{eqnarray}
where $U_i$ are $m|n$ super-matrices of generators of spin $i$. Furthermore, one can immediately identify the order of the differential operator $N$ with the parameter $N$ appearing in the OPE of $\widehat{\mathfrak{gl}(m|n)}$ currents
\begin{eqnarray}\nonumber
U_{1,ab}(z)U_{1,cd}(w)&\sim&\frac{N((-1)^{bc}\kappa \delta_{ad}\delta_{cb}+\delta_{ab}\delta_{cd})}{(z-w)^2}\\
&&+\frac{(-1)^{ab+cd+cb}\delta_{ad}U_{1,cb}-(-1)^{bc}\delta_{db}U_{1,ad}}{z-w},
\label{ref9}
\end{eqnarray}
i.e. the level of the $\widehat{\mathfrak{sl}(m|n)}_{\tilde{\kappa}}$ subalgebra equals $\tilde{\kappa}=N\kappa$. Since $N$ can be an arbitrary complex number for the general $\mathcal{W}_{m|n\times \infty}$ algebra, one might be tempted to consider a more general class of pseudo-differential operators 
\begin{eqnarray}
\mathcal{L}=(\kappa\partial)^N+U_{1}(\kappa\partial)^{N-1}+U_{2}(\kappa\partial)^{N-2}+U_{3}(\kappa\partial)^{N-3}+\dots
\end{eqnarray}
containing infinitely many super-matrix-valued fields with $U_i$ satisfying OPEs of $\mathcal{W}_{m|n\times \infty}$ with parameter $N$ equal the order of the pseudo-differential operator.

In analogy with the above analysis, one can define a coproduct on $\mathcal{W}_{m|n\times \infty}$ by multiplying two factors of this form
\begin{eqnarray}\nonumber
\mathcal{L}^{(1)}\mathcal{L}^{(2)}&=&\left ((\kappa\partial)^{N_1}+U^{(1)}_{1}(\kappa\partial)^{N_1-1}+U^{(1)}_{2}(\kappa\partial)^{N_1-2}++U^{(1)}_{3}(\kappa\partial)^{N_1-3}+\dots\right )\\
&&\times\left ((\kappa\partial)^{N_2}+U^{(2)}_{1}(\kappa\partial)^{N_2-1}+U^{(2)}_{2}(\kappa\partial)^{N_2-2}+U^{(2)}_{3}(\kappa\partial)^{N_2-3}+\dots\right )
\label{gen_miu}
\end{eqnarray}
and expressing it in terms of a pseudo-differential operator of order $N_1+N_2$ using (\ref{com_der}).

This gives an expression of the new generators $\tilde{U}_{i}$ of the algebra $\mathcal{W}_{m|n\times \infty}$ for $N=N_1+N_2$ in terms of $U^{(1)}_i$ and $U_{i}^{(2)}$ generators
\begin{eqnarray}\nonumber
\tilde{U}_{1,ij}&=&U^{(1)}_{1,ij}+U^{(2)}_{1,ij},\\ \nonumber
\tilde{U}_{2,ij}&=&U^{(1)}_{2,ij}+U^{(2)}_{2,ij}+\sum_k (U^{(1)}_{1,ik}U^{(2)}_{1,kj})+N\kappa\partial U^{(2)}_{1,kj},\\ \nonumber
\tilde{U}_{3,ij}&=&U^{(1)}_{3,ij}+U^{(2)}_{3,ij}+\sum_k (U^{(1)}_{2,ik}U^{(2)}_{1,kj})+\sum_k (U^{(1)}_{1,ik}U^{(2)}_{2,kj})\\
&&+\kappa (N-1)\sum_k (U^{(1)}_{1,ik}\partial U^{(2)}_{1,kj})+\frac{N(N-1)}{2}\kappa^2\partial^2 U^{(2)}_{1,kj}.
\end{eqnarray}
We have checked that these satisfy OPEs of $\mathcal{W}_{m|n\times \infty}$ from (\ref{allOPES}) leading us to the following conjecture:
\begin{conj}
Miura transformation from (\ref{gen_miu}) gives rise to a well-defined coproduct structure on $\mathcal{W}_{m|n\times \infty}$ with $N=N_1+N_2$ being additive.
\end{conj}

\subsection{Building blocks of truncations}
\label{blocks}

After allowing more general pseudo-differential operators in the $\mathcal{W}_{1+\infty}$ case, we have seen that there exist two more elementary Miura operators $\mathcal{L}^{(y)}_{i},\mathcal{L}^{(w)}_{i}$ expressed in terms of the generator of the $\widehat{\mathfrak{gl}(1)}$ Kac-Moody algebra associated to $y$ and $w$. Composition of such more general Miura operators yields a more general class $Y_{N_1,N_2,N_3}$ of $\mathcal{W}_{1+\infty}$ truncations. We will now use the geometric and the gauge-theoretical picture to identify the elementary algebras associated to divisors of $x,y,z,w$. These will play the role of elementary building blocks for more general truncations analogously to $\widehat{\mathfrak{gl}(1)}$ for $Y_{N_1,N_2,N_3}$. 

The identification of corresponding algebras can be done along the lines of \cite{Gaiotto:2017euk,Prochazka:2017qum}. Let us briefly review the idea behind the identification. Colored $(p,q)$-webs from figure \ref{factors} above can be given an alternative interpretation in terms of a junction of supersymmetric interfaces in a four-dimensional $\mathcal{N}=4$ super-symmetric gauge theory. Zooming out the $(p,q)$-web, we get a configuration of interfaces dividing the four-dimensional spacetime into four (or three in the special case of $n=0$ or $m=0$) corners. An analysis of corresponding interfaces \cite{Witten:2011zz,Mikhaylov:2014aoa,Gaiotto:2017euk} leads a proposal for the corresponding algebra. Interested reader should consult \cite{Gaiotto:2017euk,Prochazka:2017qum} for further details.

\paragraph{$x$-algebra}

The gauge-theoretical configuration consists of the $U(n)$ gauge theory in the upper-left corner and the $U(m)$ gauge theory in the  upper-right corner. These two are divided by an interface labeled as $(0,1)$ and both terminate on the horizontal boundary. The boundary condition on the horizontal line was identified in \cite{Gaiotto:2008sa} with the Dirichlet boundary condition and the $(0,1)$ interface couple the two theories by an additional three-dimensional bi-fundamental hypermultiplet. It was argued in \cite{Witten:2011zz,Mikhaylov:2014aoa,Gaiotto:2017euk} that such a gauge-theoretical configuration localizes to the complexified $GL(m|n)$ Chern-Simons theory on the $(0,1)$ interface with the Dirichlet boundary condition descending from the boundary conditions of the four-dimensional theory. Local operators at the boundary of the Chern-Simons theory (corner of the four-dimensional setup) are known to form the Kac-Moody algebra associated to the gauge group of the corresponding gauge theory. The level of the Kac-Moody algebra $\widehat{\mathfrak{gl}(m|n)}_\kappa$ is then given by\footnote{Note that we reproduce the relation $\Psi=\kappa+1$ from the discussion of $\mathcal{W}_{1+\infty}$ algebra if we set $m=1,n=0$.}
\begin{eqnarray}
\kappa = \Psi-m+n
\end{eqnarray}
where $\Psi$ is the coupling constant of the four-dimensional theory. The building block associated to $x$ is then the Kac-Moody algebra with OPE
\begin{eqnarray}
J_{ab}^{(x)}(z)J_{cd}^{(x)}(w)\sim  \frac{(-1)^{bc}\kappa \delta_{ad}\delta_{cb}+\delta_{ab}\delta_{cd}}{(z-w)^2}+\frac{(-1)^{ab+cd+cb}\delta_{ad}J_{cb}^{(x)}-(-1)^{bc}\delta_{cb}J_{ad}^{(x)}}{z-w} 
\label{ref8}
\end{eqnarray}
that is exactly the algebra used above in the construction of the $\mathcal{W}_{m|n\times \infty}$ algebra.

\paragraph{$y$-algebra}

The configuration associated to $y$ is analogous with only two differences. First, the orientation of the diagram is reversed that leads to the opposite sign of the level. Secondly, the vertical interface is now $(m-n,1)$ that leads to an extra shift of the level by $m-n$. The resulting algebra is the Kac-Moody algebra $\widehat{\mathfrak{gl}(m|n)}$ at level
\begin{eqnarray}
\tilde{\kappa}=-\Psi=-\kappa-m+n.
\end{eqnarray}
The generators $J^{(y)}_{ab}$ thus satisfy relations
\begin{eqnarray}\nonumber
J_{ab}^{(y)}(z)J_{cd}^{(y)}(w)&\sim&  \frac{(-1)^{bc}\tilde{\kappa} \delta_{ad}\delta_{cb}+\frac{\tilde{\kappa}}{\kappa}\delta_{ab}\delta_{cd}}{(z-w)^2}\\
&&+\frac{(-1)^{ab+cd+cb}\delta_{ad}J_{cb}^{(y)}-(-1)^{bc}\delta_{cb}J_{ad}^{(y)}}{z-w}, 
\end{eqnarray}
i.e. note the particular normalization of the diagonal $\widehat{\mathfrak{gl}(1)}$ generator.

\paragraph{$z$-algebra}

In the $z$ configuration, the path integral localizes to two copies of $GL(1)$ Chern-Simons theories at levels $\kappa$ and $-\kappa-m+n$ coupled together at the interface. It was argued in \cite{Gaiotto:2017euk,Prochazka:2017qum} that a consistent coupling requires $m$ copies of $\beta,\gamma$ ghosts (or symplectic bosons) associated to the $(1,0)$ lines ending from the right and $n$ copies of $b,c$ ghosts (free fermions) associated to $(1,0)$ lines ending from the left, all charged under the $GL(1)$ gauge group. Corresponding VOAs can be then identified with the coset
\begin{eqnarray}
\frac{\widehat{\mathfrak{gl}(1)}\times \mathcal{S}^{m|n}}{\widehat{\mathfrak{gl}(1)}}
\label{coset}
\end{eqnarray}
where $\mathcal{S}^{m|n}$ is a collection of fields $(X_i,Y_i)$ that consists of bosonic generators $\beta_i,\gamma_i$ for $i=1,\dots,m$ and fermionic generators $b_i,c_i$ for $i=m+1,\dots,m+n$ with OPEs
\begin{eqnarray}
X_i(z)Y_i(w)\sim \frac{1}{z-w},\qquad Y_i(z)X_i(w)\sim -\frac{(-1)^a}{z-w}.
\label{coset}
\end{eqnarray}
The fields $X_i$ have $+1$ under the $GL(1)$ action and $Y_i$ is are charge $-1$. More concretely, the above coset denotes the BRST cohomology of two $\widehat{\mathfrak{gl}(1)}$ currents $J,\tilde{J}$ together with  $\mathcal{S}^{m|n}$ with respect to the BRST charge
\begin{eqnarray}
Q=\oint dz \left (J-\tilde{J}+\sum_{i=1}^{m+n} (-1)^{i}(X_iY_i)\right )c
\end{eqnarray}
that imposes the above-described constraint quantum mechanically. In this expression, we have introduced an auxiliary $c,b$ ghost system. The two currents are normalized as
\begin{eqnarray}
J(z)J(w)\sim \frac{\kappa+1}{(z-w)^2},\qquad \tilde{J}(z)\tilde{J}(w)\sim \frac{-\kappa+m-n-1}{(z-w)^2}.
\end{eqnarray}

One can easily check that combinations of the form
\begin{eqnarray}
J^{(z)}_{1,ab}=(X_aY_b)+\frac{1}{\kappa+1}\delta_{ab}J
\label{Jcos}
\end{eqnarray}
are generators of the the cohomology at spin one. Moreover, combinations\footnote{For $n=m$, the expression simplifies to 
\begin{eqnarray}
U^{(z)}_{1,ab}=J^{(z)}_{1,ab}+\frac{2\kappa+1}{2\kappa(\kappa+1)}\delta_{ab}\sum_{c=1}^{m+n}(-1)^{i}J^{(z)}_{1,cc}.
\end{eqnarray}}
\begin{eqnarray}
U^{(z)}_{1,ab}=J^{(z)}_{1,ab}-\frac{1}{m-n}\left (1+\sqrt{\frac{(\kappa+1)(\kappa+m-n)}{\kappa (\kappa-m+n+1)}}\right ) \delta_{ab}\sum_{c=1}^{m+n}(-1)^{c}J^{(z)}_{1,cc}
\end{eqnarray}
generate the $\widehat{\mathfrak{gl}(m|n)}_{-1}$ Kac-Moody algebra at level
\begin{eqnarray}
\tilde{\kappa}=-1, 
\end{eqnarray}
i.e. satisfying 
\begin{eqnarray}\nonumber
U^{(z)}_{1,ab}(z)U^{(z)}_{1,cd}(w)&\sim& \frac{-(-1)^{bc} \delta_{ad}\delta_{cb}-\frac{1}{\kappa}\delta_{ab}\delta_{cd}}{(z-w)^2}\\
&&+\frac{(-1)^{ab+cd+cb}\delta_{ad}U^{(z)}_{1,cb}-(-1)^{bc}\delta_{db}U^{(z)}_{1,ad}}{z-w}.
\end{eqnarray} 

We expect the generators $J_{1,ab}^{(z)}$ to be actually strong generators of the algebra unless $(m|n)=(1|0)$ or $(2|0)$. This expectation is supported by the the following arguments. The character of (\ref{coset}) is given by the contour integral
\begin{eqnarray}
\oint_0 \frac{dz}{z}\prod_{k=0}^\infty \frac{(1+zq^{\frac{1}{2}+k})^n(1+z^{-1}q^{\frac{1}{2}+k})^n}{(1-zq^{\frac{1}{2}+k})^m(1-z^{-1}q^{\frac{1}{2}+k})^m}. 
\end{eqnarray} 
For generic values of $(m|n)$ the algebra contains less states than algebra freely generated by an $m|n$ super-matrix of free generators at spin one. One can also explicitly check this expectation at low levels for small numbers $m,n$. 

In general, normally ordered products of derivatives of $\tilde{J}_{1,ab}$ satisfy various relations that follow from the realization in terms of (\ref{Jcos}). The only exception is the $m|n=1|1$ case for which the character agrees exactly with the character of $\widehat{\mathfrak{gl}(1|1)}$. This can be understood as a consequence of the $S_4$ symmetry of the corresponding $(p,q)$-web diagram after zooming out. In such a picture, the $(p,q)$-web consists of a single $(1,0)$ and $(0,1)$ line crossing each other. All the four configurations associated to $x,y,z,w$ then correspond to the same diagram and they should all lead to the same algebra. 

Let us briefly discuss the two special examples $(m|n)=(1|0),(2|0)$. First, comparing the character of $(2|0)$ with the the character of an algebra generated by four generators at spin one, we can see that we are missing three fields at spin two. One can define the following combinations
\begin{eqnarray}
J^{(z)}_{2,ab}=(\partial X_a Y_b)-(X_a\partial Y_b) +\frac{2}{\kappa+1}(X_a(Y_bJ)),
\label{ref9}
\end{eqnarray}
and show that (at least at low levels) $J^{(z)}_{2,12},J^{(z)}_{2,21},J^{(z)}_{2,11}-J^{(z)}_{2,22}$ together with  $J^{(z)}_{1,ab}$ form the full set of strong generators in this case. These combinations $J^{(z)}_{2,ab}$ for $a\neq b$ and $J^{(z)}_{2,aa}-J^{(z)}_{2,bb}$ are in the cohomology of $Q$ also for general $m|n$ but they can be expressed in terms of fields $J^{(z)}_{1,ab}$.

For the special case $1|0$, the set of strong generators is
\begin{eqnarray}\nonumber
J^{(z)}_{1}&=&(XY)+\frac{1}{\kappa+1} J,\\ \nonumber
J^{(z)}_{2}&=&(\partial X Y)-(X\partial Y) +\frac{2}{\kappa+1} (X(YJ))+\frac{1}{(\kappa+1)^2}(JJ),\\
J^{(z)}_{3}&=&(\partial^2 X Y)-(X\partial^2 Y) +\frac{2}{\kappa+1}\partial (X(YJ))+\frac{1}{(\kappa+1)^2}(J\partial J).
\end{eqnarray}
Moreover, in analogy with (\ref{ref9}), we can define
\begin{eqnarray}
J^{(z)}_{3,ab}=(\partial^2 X_a Y_b)-(X_a\partial^2 Y_b) +\frac{2}{\kappa+1}\partial (X_a(Y_bJ))
\label{ref5}
\end{eqnarray}
and argue that combinations $J^{(z)}_{3,ab}$ for $a\neq b$ and $J^{(z)}_{3,aa}-J^{(z)}_{3,bb}$ are in the cohomology of $Q$ for all the other $m|n$ and can be expressed in terms of fields of lower spin.

Finally, note that the resulting algebra is actually independent of the parameter $\kappa$ in all the cases and one can find a more convenient description by simply sending $\kappa \rightarrow \infty$. In all the above expressions (\ref{Jcos}), (\ref{ref9}) and (\ref{ref5}), the tail of fields containing the current $J$ disappears and one finds simpler expressions
\begin{eqnarray}\nonumber
J^{(z)}_{1,ab}&=&X_a Y_b,\\
J^{(z)}_{i,ab}&=&\partial^{i-1}X_a Y_b-X_a\partial^{i-1}Y_b
\label{ref6}
\end{eqnarray} 
for $i>1$. On the other hand, we cannot simply restrict to fields at low spin in this limit since we would miss some of the generators (the simplified generators satisfy more relations). Instead of the above more-complicated expressions in terms of $X_i,Y_i$ and $J$, one can equivalently consider the algebra generated by infinitely many simpler generators of the form (\ref{ref6}). In the following sections, we will use this simpler, $\kappa$ independent realization of the algebra.

\paragraph{$w$-algebra}

The situation associated to $w$ is analogous to the $z$ case with $(X_i,Y_i)$ now being fermionic for $i=1,\dots,n$ and bosonic for $i=n+1,\dots,m+n$. Normalizing the $J,\tilde{J}$ currents as
\begin{eqnarray}
J(z)J(w)\sim -\frac{\kappa+1}{(z-w)^2},\qquad \tilde{J}(z)\tilde{J}(w)\sim \frac{\kappa+1-m+n}{(z-w)^2},
\end{eqnarray}
we have
\begin{eqnarray}
J^{(w)}_{1,ab}=-(X_aY_b)+\frac{1}{\kappa+1}\delta_{ab}J
\end{eqnarray}
generating the cohomology at spin one with the combination
\begin{eqnarray}
U^{(w)}_{1,ab}=J^{(z)}_{1,ab}-\frac{1}{m-n}\left (1+\sqrt{\frac{(\kappa+1)(\kappa+m-n)}{\kappa (\kappa-m+n+1)}}\right ) \delta_{ab}\sum_{c=1}^{m+n}(-1)^{c}J^{(w)}_{cc}
\end{eqnarray}
identified with the standard generators of the $\widehat{\mathfrak{gl}(m|n)}_1$ Kac-Moody algebra at level
\begin{eqnarray}
\tilde{\kappa}=1,
\end{eqnarray}
i.e. satisfying OPE
\begin{eqnarray}\nonumber
U_{1,ab}^{(w)}(z)U_{cd}^{(w)}(w)&\sim& \frac{(-1)^{bc} \delta_{ad}\delta_{cb}+\frac{1}{\kappa}\delta_{ab}\delta_{cd}}{(z-w)^2}\\
&&+\frac{(-1)^{ab+cd+cb}\delta_{ad}J_{1,cb}^{(i)}-(-1)^{bc}\delta_{db}J_{1,ad}^{(w)}}{z-w}.
\end{eqnarray} 
The discussion of strong generators is the same as above just with the role of $m$ and $n$ interchanged and up to some minus signs. For example at spin two, we have the following combination
\begin{eqnarray}
J^{(w)}_{2,ab}=-\partial X_a Y_b+X_a\partial Y_b +\frac{2}{\kappa+1}X_aY_bJ.
\end{eqnarray}
Instead of these generators, we will again consider the realization of the algebra coming from the $\kappa \rightarrow \infty$ limit, that can be realized in terms of
\begin{eqnarray}\nonumber
J^{(w)}_{1,ab}&=&X_a Y_b,\\
J^{(w)}_{i,ab}&=&\partial^{i-1}X_a Y_b-X_a\partial^{i-1}Y_b.
\label{ref4}
\end{eqnarray} 
as above.

\subsection{$x^{N_3} y^{N_2} z^{N_4} w^{N_1}$-algebras}
\label{composition}

According to the above analysis, the elementary algebras $x,y,z,w$ form particular truncations of the $\mathcal{W}_{m|n\times \infty}$ algebra. We will now determine the specialization of the parameter $N$ associated to such elementary truncations and find combinations of their fields that satisfy OPEs of $\mathcal{W}_{m|n\times \infty}$ in the $U$-basis with correctly specialized parameter $N$. Using such an explicit realization of $U$-generators and the coproduct from the above section, one can define a general $x^{N_3} y^{N_2} z^{N_4} w^{N_1}$-algebra by multiplying the elementary Miura factors associated to $x,y,z,w$.

Since $N$ is additive under the coproduct, we can immediately identify its specialization associated to the general $x^{N_3} y^{N_2} z^{N_4} w^{N_1}$-algebra by determining $N$ of the elementary factors. As discussed bellow, one recovers
\begin{eqnarray}
N=\frac{((n-m)h_1-h_2)N_3+h_2N_2 -h_1N_4+h_1N_1}{(n-m)h_1-h_2}
\label{order}
\end{eqnarray}
that is exactly (up to the overall normalization) the charge of $x^{N_3} y^{N_2} z^{N_4} w^{N_1}$ from (\ref{charge}). The normalization is such that the order of $\mathcal{L}_i^{(x)}$ is $1$. For $m|n=1|0$, the above expression simplifies to
\begin{eqnarray}
N=\frac{h_3 N_3 +h_2 N_2 +h_1N_1}{h_3}
\label{Cspec}
\end{eqnarray}
from \cite{Prochazka:2017qum}.

\paragraph{$x$-algebra}

The case of the $x$-algebra is trivial. Comparing (\ref{ref8}) with the general expression (\ref{ref9}), one gets order $N_x=1$ as expected. The corresponding Miura operator is simply
\begin{eqnarray}
\mathcal{L}^{(x)}_i=\kappa \partial + J^{(x)}_{i}.
\end{eqnarray}

\paragraph{$y$-algebra}

We will now look at the $y$-algebra as a quotient of the two-parameter family of algebras (\ref{allOPES}) parametrized by $\kappa=\Psi-m+n$ and $N$. As discussed above, the $y$-algebra contains the Kac-Moody algebra $\widehat{\mathfrak{sl}(m|n)}_{\tilde{\kappa}}$ at level $\tilde{\kappa}=-\Psi=-\kappa-m+n$. One can then write
\begin{eqnarray}
\tilde{\kappa}=-\Psi=-\Psi\frac{\Psi-m+n}{\Psi-m+n}=\frac{h_2}{(n-m)h_1-h_2}\kappa=N_y\kappa.
\end{eqnarray}
and identify the specialization $N_y$ of the parameter $N$ consistent with the formula (\ref{order}) for $N_2=1, N_1=N_3=N_4=0$.

It is straightforward to identify the first few generators $U^{(y)}_{i,ab}$ satisfying (\ref{allOPES}) with $N$ correctly specialized in terms of the Kac-Moody generators $J^{(y)}_{ab}$. One gets
\begin{eqnarray}\nonumber
U^{(y)}_{1,ab}&=&J^{(y)}_{ab}\\ \nonumber
U^{(y)}_{2,ab}&=&(J^{(y)}J^{(y)})_{ab}-\frac{\mbox{str}(J^{(y)})J^{(y)}_{ab}}{\kappa+m-n}-\frac{\delta_{ab}\left (\mbox{str}(J^{(y)})\right )^2}{2\kappa(\kappa+m-n)}+\frac{\delta_{ab}\mbox{str} (J^{(y)}J^{(y)})}{2\kappa}\\ 
&&-\kappa \partial J^{(y)}_{ab}-\frac{1}{2}\delta_{ab}\mbox{str}(\partial J^{(y)}).
\end{eqnarray} 
where we have introduced notation
\begin{eqnarray}
\mbox{str}(A)=\sum_{c=1}^{n+m}(-1)^{c}A_{cc},\qquad (AA)_{ab}= \sum_{c=1}^{n+m}A_{ac}A_{cb}
\end{eqnarray}
and all the products are normally ordered. Note that an existence of the realization of $U^{(y)}_{2,ab}$ in terms of $J^{(y)}_{ab}$ is a non-trivial test of our proposal for generalized the Miura transformation for truncations of $\mathcal{W}_{m|n\times \infty}$ beyond $x^N$-algebras.

\paragraph{$z$-algebra}

Similarly as above, we have now an algebra containing   the Kac-Moody algebra $\widehat{\mathfrak{sl}(m|n)}_{-1}$ at level $\tilde{\kappa}=-1$ and we can again write
\begin{eqnarray}
\tilde{\kappa}=-1=-\frac{\Psi+m-n}{\Psi+m-n}=-\frac{h_1}{(m-n)h_1-h_2}\kappa=N\kappa
\end{eqnarray}
and identify the specialization $N_z$ of the parameter $N$ consistent with the formula (\ref{order}) for $N_4=1, N_1=N_2=N_3=0$.

It is straightforward to identify the first few generators $U^{(z)}_{i,ab}$ satisfying (\ref{allOPES}) with $N$ correctly specialized in terms of the of the fields $J^{(z)}_{1,ab}=(X_aY_b)$ or possibly higher-spin generators $J^{(z)}_{i,ab}=(\partial^{i-1} X_{a}Y_b)-(X_{a}\partial^{i-1} Y_b)$ for $i>1$. One finds
\begin{eqnarray} \nonumber
U^{(z)}_{1,ab}&=&J^{(z)}_{1,ab}+\frac{1}{\kappa-\hat{\kappa}}\delta_{ab}\ \mbox{str} (J^{(z)}_1),\\ \nonumber
U^{(z)}_{2,ab}&=&\frac{\kappa+1}{2(m-n)\kappa}\bigg ( (2\kappa-(m-n-2)\hat{\kappa})J^{(z)}_{2,ab}-(\kappa+\hat{\kappa})\delta_{ab}\ \mbox{str}(J^{(z)}_{2})\\ \nonumber
&&-2(\kappa+\hat{\kappa})(J^{(z)}_1J^{(z)}_1)_{ab}+\left (1+2\frac{\kappa+\tilde{\kappa}}{m-n}\right )\delta_{ab}\left (\mbox{str} (J^{(z)}_1)\right )^2\\
&&+(2\kappa+\hat{\kappa})\partial J^{(z)}_{1,ab}+2(\kappa+\hat{\kappa})\delta_{ab}\ \mbox{str}(\partial J^{(z)}_1) \bigg ).
\end{eqnarray} 
where we have introduced a constant
\begin{eqnarray}
\hat{\kappa}=\sqrt{\kappa (\kappa+m-n)}.
\end{eqnarray}
For $m=n$, the above expression simplifies to
\begin{eqnarray} 
U^{(z)}_{1,ab}&=&J^{(z)}_{1,ab}+\frac{1}{2\kappa}\delta_{ab}\ \mbox{str} (J^{(z)}_1),\\ \nonumber
U^{(z)}_{2,ab}&=&\frac{\kappa+1}{2\kappa}\bigg ((\kappa-1)J^{(z)}_{2,ab}+\frac{1}{2}\delta_{ab}\ \mbox{str}(J^{(z)}_{2})+(J^{(z)}_1J^{(z)}_1)_{ab}\\ \nonumber
&&+\frac{1}{4\kappa}\delta_{ab}\left (\mbox{str} (J^{(z)}_1)\right )^2-\kappa \partial J^{(z)}_{1,ab}-\delta_{ab}\ \mbox{str}(\partial J^{(z)}_1) \bigg ).
\end{eqnarray} 

\paragraph{$w$-algebra}

The identification of the parameter $N$ is the same as in the above case of the $z$-algebra up to an overall minus sign consistently with (\ref{order}). Similarly, the expression for $U^{(w)}_{i,ab}$ has the same form
\begin{eqnarray} 
U^{(w)}_{1,ab}&=&-J^{(w)}_{1,ab}-\frac{1}{\kappa-\hat{\kappa}}\delta_{ab}\ \mbox{str} (J^{(w)}_1)\\ \nonumber
U^{(w)}_{2,ab}&=&-\frac{\kappa-1}{2(m-n)\hat{\kappa}}\bigg ( (2\hat{\kappa}-(n-m-2)(\kappa+m-n))J^{(w)}_{2,ab}\\ \nonumber
&&-(\kappa+\hat{\kappa}+m-n)\left (\delta_{ab}\ \mbox{str}(J^{(w)}_{2})+2(J^{(w)}_1J^{(w)}_1)_{ab}-2\delta_{ab}\ \mbox{str}(\partial J^{(w)}_1)  \right )\\  \nonumber
&&-\frac{\hat{\kappa}}{\kappa}\left (1+2\frac{\kappa+\tilde{\kappa}}{m-n}\right )\delta_{ab}\left (\mbox{str} (J^{(w)}_1)\right )^2-(m-n)(2\hat{\kappa}+\kappa+m-n)\partial J^{(w)}_{1,ab}\bigg ).
\end{eqnarray} 
up to the exchange of $\kappa \leftrightarrow \hat{\kappa}$ together with some shifts by factors of $m-n$ and some minus signs. For $m=n$, this expression simplifies to
\begin{eqnarray} 
U^{(w)}_{1,ab}&=&-J^{(w)}_{1,ab}-\frac{1}{2\kappa}\delta_{ab}\ \mbox{str} (J^{(w)}_1)\\ \nonumber
U^{(w)}_{2,ab}&=&-\frac{\kappa-1}{2\kappa}\bigg ((\kappa+1)J^{(w)}_{2,ab}+\frac{1}{2}\delta_{ab}\ \mbox{str}(J^{(w)}_{2})-(J^{(w)}_1J^{(w)}_1)_{ab}\\ \nonumber
&&+\frac{1}{4\kappa}\delta_{ab}\left (\mbox{str} (J^{(w)}_1)\right )^2-\kappa \partial J^{(w)}_{1,ab}-\delta_{ab}\ \mbox{str}(\partial J^{(w)}_1) \bigg ).
\end{eqnarray}

Having determined the elementary Miura factors $\mathcal{L}^{(x)}$, $\mathcal{L}^{(y)}$, $\mathcal{L}^{(z)}$, $\mathcal{L}^{(w)}$, one can define an algebra associated to any function of the form $x^{N_3} y^{N_2} z^{N_4} w^{N_1}$ determined as a subalgebra inside a tensor product of $N_3$ copies of the $\widehat{\mathfrak{gl}(m|n)}_{\kappa}$ Kac-Moody algebras, $N_2$ copies of the $\widehat{\mathfrak{gl}(m|n)}_{-\kappa-m+n}$ Kac-Moody algebras, $N_4$ copies of $X_iY_i$ containing $m$ pairs of bosonic fields and $n$ pairs of fermionic fiels, and finally $N_1$ copies of $X_iY_i$ containing $m$ pair of fermionic fields and $n$ pairs of bosonic fiels. The subalgebra can be obtained similarly as above in the $\mathcal{W}_{1+\infty}$ case or the $x^{N_3}$-algebras from the derivation of the OPEs of $\mathcal{W}_{m|n\times \infty}$ by multiplying
\begin{eqnarray}
\mathcal{L}^{(x)}_1\dots \mathcal{L}^{(x)}_{N_3}\mathcal{L}^{(y)}_{N_3+1}\dots \mathcal{L}^{(y)}_{N_3+N_2}\mathcal{L}^{(z)}_{N_3+N_2+1}\dots \mathcal{L}^{(z)}_{N_3+N_2+N_4}\mathcal{L}^{(w)}_{N_3+N_2+N_4+1}\dots \mathcal{L}^{(w)}_{N_3+N_2+N_4+N_1}.
\label{ref10}
\end{eqnarray}
and rewriting it as a matrix-valued pseudo-differential operator in the standard form
\begin{eqnarray}
=(\kappa\partial)^{N}+U_{1}(\kappa\partial)^{N-1}+U_{2}(\kappa\partial)^{N-2}+\dots
\end{eqnarray}
with $U_i$ super-matrices of generators at spin $i$ and $\mathcal{L}^{(\alpha)}$ being the following pseudo-differential operator
\begin{eqnarray}
\mathcal{L}^{(\alpha)}=(\kappa\partial)^{N_\alpha}+U_{1}^{(\alpha)}(\kappa\partial)^{N_\alpha-1}+U_{2}^{(\alpha)}(\kappa\partial)^{N_\alpha-2}+\dots
\end{eqnarray}
for $\alpha=x,y,z,w$. This gives in particular the following expression for the spin-one and the spin-two fields
\begin{eqnarray}\nonumber
U_{1,ab}&=&\sum_{i=1}^{N_1+N_2+N_3+N_4}U^{(\alpha_i)}_{1,ab},\\ \nonumber
U_{2,ab}&=&\sum_{i=1}^{N_1+N_2+N_3+N_4}U^{(\alpha_i)}_{2,ab}+\sum_{i=1}^{N_1+N_2+N_3+N_4}\sum_{j=i+1}^{N_1+N_2+N_3+N_4}(U^{(\alpha_i)}_{1}U^{(\alpha_j)}_{1})_{ab}\\
&&+\kappa \sum_{i=1}^{N_1+N_2+N_3+N_4}\sum_{j=1}^{i-1} N_{\alpha_j}\partial U^{(\alpha_i)}_{1,ab} .
\label{coproduct2}
\end{eqnarray}

Let us conclude by stating the following conjecture:
\begin{conj}
The $U_i$ generators of $\mathcal{W}_{m|n\times \infty}$ with the parameter $N$ specialized to $N_\alpha$ can be realized in terms of the elementary algebra $\alpha=x,y,z,w$. Coproduct of $\mathcal{W}_{m|n\times \infty}$ then gives rise to a free-field realization of the $x^{N_3}y^{N_2}z^{N_4}w^{N_1}$ algebra.
\label{conj2}
\end{conj}

\subsection{Bosonization-like relations}
\label{bosonization}

Remember the relation $xy=z^mw^n$ in the ring of holomorphic functions on $CY^{3}_{m,n}$ that relate functions $x^{N_3-1}y^{N_2-1}z^{N_4}w^{N_1}$ and  $x^{N_3}y^{N_2}z^{N_4+m}w^{N_1+n}$. On the other hand both $x^{N_3-1}y^{N_2-1}z^{N_4}w^{N_1}$ and  $x^{N_3}y^{N_2}z^{N_4+m}w^{N_1+n}$ corresponds to a different free-field realization using the Miura transformation (\ref{ref10}). As a consequence of the above discussion, the generators of the algebra satisfy the same relations but one algebra might still be a subalgebra of the second other. Since both functions correspond to the same physical setup, we actually expect this not to be the case and the algebras should be equal on the nose. This leads to the following conjecture:
\begin{conj}
The pseudo-differential operators for $CY^{3}_{m,n}$ satisfy
\begin{equation}
\mathcal{L}^{(x)}\mathcal{L}^{(y)}=\left (\mathcal{L}^{(z)}\right )^m\left (\mathcal{L}^{(w)}\right )^n,
\end{equation}
i.e. algebras on both sides have the same size and satisfy the same OPEs.
\end{conj}

Let us illustrate this relation on the simplest example of $xy=z$ for $CY^3_{1,0}$. In particular, we will now show that the free-field realizations of fields $U_1,U_2$ can be related by the standard bosonization of $\beta,\gamma$ ghosts. The two Miura operators associated to $x$ and $y$ are
\begin{eqnarray}
\mathcal{L}^{(x)}&=&\kappa \partial + J^{(x)}\\ \nonumber
\mathcal{L}^{(y)}&=&(\kappa \partial)^{-\frac{1}{\kappa}-1} + J^{(y)}(\kappa \partial)^{-\frac{1}{\kappa}-2}+\frac{1+2\kappa}{2}\left (\frac{J^{(y)}J^{(y)}}{1+\kappa}-\partial J^{(y)} \right )(\kappa \partial)^{-\frac{1}{\kappa}-3}+\dots,
\end{eqnarray}
with normalization
\begin{eqnarray}
J^{(x)}(z)J^{(x)}(w)\sim \frac{\kappa+1}{(z-w)^2},\qquad J^{(y)}(z)J^{(y)}(w)\sim -\frac{(\kappa+1)^2}{\kappa}\frac{1}{(z-w)^2}.
\end{eqnarray}
Multiplying these two $\mathcal{L}^{(y)}\mathcal{L}^{(x)}$ gives
\begin{eqnarray}
\bigg [1 +(J^{(x)}+ J^{(y)})(\kappa \partial)^{-1}&+&\bigg (\frac{1+2\kappa}{2}\bigg (\frac{J^{(y)}J^{(y)}}{1+\kappa}-\partial J^{(y)}\bigg )\\ \nonumber
&&+ J^{(x)}J^{(y)}-\kappa \bigg (1+\frac{1}{\kappa}\bigg )\partial J^{(x)} \bigg )(\kappa \partial)^{-2}+\dots\bigg ](\kappa \partial)^{-\frac{1}{\kappa}}.
\end{eqnarray}
On the other hand, we have for the free-field realization of the $z$-algebra in terms of the $\beta,\gamma$ system
\begin{eqnarray}
U^{(z)}_1&=&J_1^{(z)}\\ \nonumber
U^{(z)}_2&=&\frac{1+\kappa}{2}\left (J_2^{(z)}+\frac{1}{\kappa}J^{(z)}_1J^{(z)}_1-\frac{\sqrt{1+\kappa}}{\sqrt{\kappa}}\partial J^{(z)}_1 \right ) ,
\end{eqnarray}
where $J^{(z)}_1=\beta \gamma$ and $J^{(z)}_2=\partial \beta \gamma-\beta \partial \gamma$.

We can bosonize the $\beta,\gamma$ system by introducing
\begin{eqnarray}
\beta=e^{\phi-\chi},\qquad \gamma =\partial \chi e^{-\phi+\chi}
\end{eqnarray}
where $\phi$ and $\chi$ satisfy
\begin{eqnarray}
\phi (z)\phi (w)\sim -\log (z-w),\qquad \chi (z)\chi (w)\sim \log (z-w).
\end{eqnarray}
A simple calculation leads us to the following identification
\begin{eqnarray}\nonumber
J^{(z)}_1&=&\partial \phi \\
J^{(z)}_2&=& (\partial \phi )^2-(\partial \chi )^2-\partial^2 \chi
\end{eqnarray}
that are expressions simply in terms of $\partial \phi$ and $\partial \chi$ that satisfy OPEs of $\widehat{\mathfrak{gl}(1)}$ currents.  It is then easy to check that transformation
\begin{eqnarray}\nonumber
\partial \phi &=& -\sqrt{\frac{\kappa}{\kappa+1}}\left (J^{(x)}+J^{(y)} \right )\\
\partial \chi &=& J^{(x)}+\frac{\kappa}{\kappa+1}J^{(y)}
\end{eqnarray}
relates the two free-field realizations and can be thought of as the VOA analogue of the relation $xy=z$.

The general situation is much more complicated but might likely be proved by the use of the Wakimoto realization together with the bosonization of $\beta,\gamma$ and $b,c$ systems. The proper analysis is left for future work but let us at least count the number of fields on both sides of such a generalized bosonization corresponding to $xy=z^mw^n$. 

First, the algebra $\widehat{\mathfrak{gl}(m|n)}$ admits the Wakimoto free-field realization \cite{Wakimoto:1986gf,Feigin:1990qn,Yang:2007zzb} in terms of
\begin{itemize}
\item $(m+n)$ free bosons associated to Cartan genrators,
\item $\frac{m(m-1)}{2}+\frac{n(n-1)}{2}$ copies of $(\beta,\gamma)$ ghosts associated to positive bosonic roots,
\item $mn$ copies of $(b,c)$ ghosts associated to positive fermionic roots.
\end{itemize} 
The left hand side associated to $xy$ can be thus realized in terms of two copies of the above-listed fields. 

On the other hand, the right hand side corresponding to $z^mw^n$ is realized by $m$ copies of
\begin{itemize}
\item $m$ copies of $(\beta,\gamma)$ ghosts,
\item $n$ copies $(b,c)$ ghosts,
\end{itemize} 
together with $n$ copies of
\begin{itemize}
\item $m$ copies of $(b,c)$ ghosts,
\item $n$ copies of $(\beta,\gamma)$ ghosts.
\end{itemize} 
Putting everything together, we have an algebra realized in terms of $m^2+n^2$ copies of the $(\beta,\gamma)$ system and $2mn$ copies of $(b,c)$ ghosts. Comparing with twice the Wakimoto realization above, we see that we have a mismatch of $n+m$ copies of $(\beta,\gamma)$ ghosts on one side and $2(m+n)$ free bosons on the other side. It is likely that these $(\beta,\gamma)$ systems can be bosonized and the two free-field realizations related.

\section{Conclusion}

Let us finish by stating the main results of the above discussion:
\begin{enumerate}
\item We find OPEs of spin-one and spin-two generators of a two-parameter family of algebras denoted as $\mathcal{W}_{m|n \times \infty}$ using a matrix generalization of the Miura transformation. We defined a coproduct structure on the algebra using general pseudo-differential matrix-valued Miura operators. 
\item We identify a large class of truncations of $\mathcal{W}_{m|n\times \infty}$ associated to a specialization of the parameter $N$ of the algebra
\begin{eqnarray}
N=\frac{((n-m)h_1-h_2)N_3+h_2N_2-h_1N_4+h_1N_1}{(n-m)h_1-h_2}.
\end{eqnarray}
\item Using gauge-theoretical considerations, we identify VOAs associated to the four elementary truncations labeled as $x,y,z,w$. In particular, we identify them with $\widehat{\mathfrak{gl}(m|n)}$ Kac-Moody algebras together with subalgebras of various copies of $(\beta,\gamma)$ and $(b,c)$ systems. We  show that $U_{1}$ and $U_2$ generators of $\mathcal{W}_{m|n\times \infty}$ can be realized in terms of these algebras if we correctly specialize the parameter $N$. 
\item We propose that general truncations associated to general functions $ x^{N_3} y^{N_2} z^{N_4} w^{N_1}$ can be obtained by composing various matrix-valued pseudo-differential operators $\mathcal{L}^{(\alpha)}$ for $\alpha=x,y,z,w$.
\item We conjecture new bosonisation-like relations between different free-field realizations inherited from the relation $xy=z^mw^n$.
\item We find a non-trivial interplay between properties of $\mathcal{W}_{m|n\times \infty}$ and its truncations and the geometry of $CY^{3}_{m,n}$ singularities. Such a correspondence can be schematically summarized as:
\begin{center}
\begin{tabular}{ | c | c| } 
\hline
Geometry & Algebra\\ 
\hline
$CY^{3}_{m,n}: xy=z^mw^n$ & $\mathcal{W}_{m|n\times \infty }$ algebra \\ 
Elementary function $x$ & $\widehat{\mathfrak{gl}(m|n)}_{\kappa}$ with Miura operator $\mathcal{L}^{(x)}$  \\ 
Elementary function $y$ & $\widehat{\mathfrak{gl}(m|n)}_{-\kappa-m+n}$ with Miura operator $\mathcal{L}^{(y)}$  \\ 
Elementary function $z$ & $\widehat{\mathfrak{gl}(m|n)}_{-1}$ with Miura operator $\mathcal{L}^{(z)}$  \\ 
Elementary function $w$ & $\widehat{\mathfrak{gl}(m|n)}_{1}$ with Miura operator $\mathcal{L}^{(w)}$  \\ 
Function $x^{N_3} y^{N_2} z^{N_4} w^{N_1}$ & Truncation $( \mathcal{L}^{(x)})^{N_3} (\mathcal{L}^{(y)})^{N_2}( \mathcal{L}^{(z)} )^{N_4}( \mathcal{L}^{(w)})^{N_1}$  \\ 
Relation $xy=z^mw^n$ & Bosonization-like relations  \\ 
Equivariant parameters $h_1,h_2$ & Parameter $\Psi =-\frac{h_2}{h_1}$  \\ 
Charge of $x^{N_3} y^{N_2} z^{N_4} w^{N_1}$ & Specialization of parameter $N$  \\ 
\hline
\end{tabular}
\end{center}

\end{enumerate}

Apart from the rigorous proof of the above-stated conjectures 1,2, 3 and 4, the above analysis raises many questions. Let us mention at least some of them. In \cite{Prochazka:2019dvu}, a simple realization of the elementary Miura operators for $\mathcal{W}_{1+\infty}$ was found. Is there such a closed-form expression also for the more general  $\mathcal{W}_{m|n\times \infty}$ algebras? What is the physical origin of the Miura operators? The original motivation for the project was understanding VOAs associated to a large class of generalizations of the AGT correspondence. Can we proof such a correspondence beyond the $\mathcal{W}_{1+\infty}$ case? Can we extend the Miura transformation beyond $\mathcal{W}_{m|n\times \infty}$ and construct truncations of shifted versions of $\mathcal{W}_{m|n\times \infty}$? Can we extend the above analysis to more complicated geometries leading for example to a $D(2,1;-\Psi)_N$ version of $\mathcal{W}_{1+\infty}$. Note that in this case, both parameters of the infinite algebra are visible already at level one. One should be able to identify specializations of $N$ by an analysis of null states of its $\widehat{D(2,1;-\Psi)}_N$ subalgebra.

\section*{Acknowledgements} 
I  am particularly grateful to Davide Gaiotto for many suggestions and Tom\'{a}\v{s} Proch\'{a}zka for sharing his unpublished results that triggered my interest in the topic. I  would like to thank Mina Aganagic, Kevin Costello, Thomas Creutzig, Tadashi Okazaki, Yan Soibelman, Yaping Yang, Gufang Zhao and Yehao Zhou for discussions on related topics. I am thankful to Kris Thielemans for his Mathematica package OPEdefs. The research of MR was supported by NSF grant 1521446, NSF grant 1820912, the Berkeley Center for Theoretical Physics, the Simons Foundation and Perimeter Institute for Theoretical Physics. Research at Perimeter Institute is supported by the Government of Canada through the Department of Innovation, Science and Economic Development and by the Province of Ontario through the Ministry of Research, \& Innovation and Science.

\appendix

\bibliography{miura}

\providecommand{\href}[2]{#2}\begingroup\raggedright\begin{thebibliography}{10}

\bibitem{Zamolodchikov:1985wn}
A.~B. Zamolodchikov, {\it {Infinite Additional Symmetries in Two-Dimensional
  Conformal Quantum Field Theory}},  {\em Theor. Math. Phys.} {\bf 65} (1985)
  1205--1213. [Teor. Mat. Fiz.65,347(1985)].

\bibitem{Fateev:1987zh}
V.~A. Fateev and S.~L. Lukyanov, {\it {The Models of Two-Dimensional Conformal
  Quantum Field Theory with Z(n) Symmetry}},  {\em Int. J. Mod. Phys.} {\bf A3}
  (1988) 507. [,507(1987)].

\bibitem{Drinfeld:1984qv}
V.~G. Drinfeld and V.~V. Sokolov, {\it {Lie algebras and equations of
  Korteweg-de Vries type}},  {\em J. Sov. Math.} {\bf 30} (1984) 1975--2036.

\bibitem{Bais:1987dc}
F.~A. Bais, P.~Bouwknegt, M.~Surridge, and K.~Schoutens, {\it {Extensions of
  the Virasoro Algebra Constructed from Kac-Moody Algebras Using Higher Order
  Casimir Invariants}},  {\em Nucl. Phys.} {\bf B304} (1988) 348--370.

\bibitem{Goddard:1984vk}
P.~Goddard, A.~Kent, and D.~I. Olive, {\it {Virasoro Algebras and Coset Space
  Models}},  {\em Phys. Lett.} {\bf 152B} (1985) 88--92.

\bibitem{Goddard:1986ee}
P.~Goddard, A.~Kent, and D.~I. Olive, {\it {Unitary Representations of the
  Virasoro and Supervirasoro Algebras}},  {\em Commun. Math. Phys.} {\bf 103}
  (1986) 105--119.

\bibitem{Bais:1987zk}
F.~A. Bais, P.~Bouwknegt, M.~Surridge, and K.~Schoutens, {\it {Coset
  Construction for Extended Virasoro Algebras}},  {\em Nucl. Phys.} {\bf B304}
  (1988) 371--391.

\bibitem{Bowcock:1990ku}
P.~Bowcock, {\it {Quasi-primary Fields and Associativity of Chiral Algebras}},
  {\em Nucl. Phys.} {\bf B356} (1991) 367--386.

\bibitem{Kausch:1990bn}
H.~G. Kausch and G.~M.~T. Watts, {\it {A Study of W algebras using Jacobi
  identities}},  {\em Nucl. Phys.} {\bf B354} (1991) 740--768.

\bibitem{Bershadsky:1989mf}
M.~Bershadsky and H.~Ooguri, {\it {Hidden SL(n) Symmetry in Conformal Field
  Theories}},  {\em Commun. Math. Phys.} {\bf 126} (1989) 49. [,49(1989)].

\bibitem{Feigin:1990pn}
B.~Feigin and E.~Frenkel, {\it {Quantization of the Drinfeld-Sokolov
  reduction}},  {\em Phys. Lett.} {\bf B246} (1990) 75--81.

\bibitem{Alday:2009aq}
L.~F. Alday, D.~Gaiotto, and Y.~Tachikawa, {\it {Liouville Correlation
  Functions from Four-dimensional Gauge Theories}},  {\em Lett. Math. Phys.}
  {\bf 91} (2010) 167--197, [\href{http://arxiv.org/abs/0906.3219}{{\tt
  arXiv:0906.3219}}].

\bibitem{Wyllard:2009hg}
N.~Wyllard, {\it {A(N-1) conformal Toda field theory correlation functions from
  conformal N = 2 SU(N) quiver gauge theories}},  {\em JHEP} {\bf 11} (2009)
  002, [\href{http://arxiv.org/abs/0907.2189}{{\tt arXiv:0907.2189}}].

\bibitem{Nekrasov:2002qd}
N.~A. Nekrasov, {\it {Seiberg-Witten prepotential from instanton counting}},
  {\em Adv. Theor. Math. Phys.} {\bf 7} (2003), no.~5 831--864,
  [\href{http://arxiv.org/abs/hep-th/0206161}{{\tt hep-th/0206161}}].

\bibitem{Nakajima:1994nid}
H.~Nakajima, {\it {Instantons on ALE spaces, quiver varieties, and Kac-Moody
  algebras}},  {\em Duke Math. J.} {\bf 76} (1994), no.~2 365--416.

\bibitem{Schiffmann:2012gf}
O.~Schiffmann and E.~Vasserot, {\it {Cherednik algebras, W algebras and the
  equivariant cohomology of the moduli space of instantons on A2}},
  \href{http://arxiv.org/abs/1202.2756}{{\tt arXiv:1202.2756}}.

\bibitem{Maulik:2012wi}
D.~Maulik and A.~Okounkov, {\it {Quantum Groups and Quantum Cohomology}},
  \href{http://arxiv.org/abs/1211.1287}{{\tt arXiv:1211.1287}}.

\bibitem{Braverman:2014xca}
A.~Braverman, M.~Finkelberg, and H.~Nakajima, {\it {Instanton moduli spaces and
  $\mathcal W$-algebras}},  \href{http://arxiv.org/abs/1406.2381}{{\tt
  arXiv:1406.2381}}.

\bibitem{Yu:1991bk}
F.~Yu and Y.-S. Wu, {\it {Nonlinearly deformed W(infinity) algebra and second
  Hamiltonian structure of KP hierarchy}},  {\em Nucl. Phys.} {\bf B373} (1992)
  713--734.

\bibitem{deBoer:1993gd}
J.~de~Boer, L.~Feher, and A.~Honecker, {\it {A Class of W algebras with
  infinitely generated classical limit}},  {\em Nucl. Phys.} {\bf B420} (1994)
  409--446, [\href{http://arxiv.org/abs/hep-th/9312049}{{\tt hep-th/9312049}}].
  [,409(1993)].

\bibitem{Khesin:1994ey}
B.~Khesin and F.~Malikov, {\it {Universal Drinfeld-Sokolov reduction and
  matrices of complex size}},  {\em Commun. Math. Phys.} {\bf 175} (1996)
  113--134, [\href{http://arxiv.org/abs/hep-th/9405116}{{\tt hep-th/9405116}}].

\bibitem{Hornfeck:1994is}
K.~Hornfeck, {\it {W algebras of negative rank}},  {\em Phys. Lett.} {\bf B343}
  (1995) 94--102, [\href{http://arxiv.org/abs/hep-th/9410013}{{\tt
  hep-th/9410013}}].

\bibitem{Blumenhagen:1994wg}
R.~Blumenhagen, W.~Eholzer, A.~Honecker, K.~Hornfeck, and R.~Hubel, {\it {Coset
  realization of unifying W algebras}},  {\em Int. J. Mod. Phys.} {\bf A10}
  (1995) 2367--2430, [\href{http://arxiv.org/abs/hep-th/9406203}{{\tt
  hep-th/9406203}}].

\bibitem{Gaberdiel:2012aa}
M.~R. Gaberdiel and R.~Gopakumar, {\it Triality in minimal model holography},
  \href{http://arxiv.org/abs/1205.2472}{{\tt arXiv:1205.2472}}.

\bibitem{Prochazka:2014aa}
T.~Proch\'azka, {\it Exploring $\mathcal{W}_{\infty}$ in the quadratic basis},
  \href{http://arxiv.org/abs/1411.7697}{{\tt arXiv:1411.7697}}.

\bibitem{Linshaw:2017tvv}
A.~R. Linshaw, {\it {Universal two-parameter $\mathcal{W}_{\infty}$-algebra and
  vertex algebras of type $\mathcal{W}(2,3,\dots, N)$}},
  \href{http://arxiv.org/abs/1710.02275}{{\tt arXiv:1710.02275}}.

\bibitem{Prochazka:2017qum}
T.~Procházka and M.~Rapčák, {\it {Webs of W-algebras}},  {\em JHEP} {\bf 11}
  (2018) 109, [\href{http://arxiv.org/abs/1711.06888}{{\tt arXiv:1711.06888}}].

\bibitem{Prochazka:2015deb}
T.~Procházka, {\it {$ \mathcal{W}$-symmetry, topological vertex and affine
  Yangian}},  {\em JHEP} {\bf 10} (2016) 077,
  [\href{http://arxiv.org/abs/1512.07178}{{\tt arXiv:1512.07178}}].

\bibitem{Bershtein}
M.~Bershtein, B.~L. Feigin, and G.~Merzon, {\it Plane partitions with a "pit":
  generating functions and representation theory},
  \href{http://arxiv.org/abs/1512.08779}{{\tt arXiv:1512.08779}}.

\bibitem{Litvinov:2016mgi}
A.~Litvinov and L.~Spodyneiko, {\it {On W algebras commuting with a set of
  screenings}},  {\em JHEP} {\bf 11} (2016) 138,
  [\href{http://arxiv.org/abs/1609.06271}{{\tt arXiv:1609.06271}}].

\bibitem{Gaiotto:2017euk}
D.~Gaiotto and M.~Rapčák, {\it {Vertex Algebras at the Corner}},  {\em JHEP}
  {\bf 01} (2019) 160, [\href{http://arxiv.org/abs/1703.00982}{{\tt
  arXiv:1703.00982}}].

\bibitem{Leung:1997tw}
N.~C. Leung and C.~Vafa, {\it {Branes and toric geometry}},  {\em Adv. Theor.
  Math. Phys.} {\bf 2} (1998) 91--118,
  [\href{http://arxiv.org/abs/hep-th/9711013}{{\tt hep-th/9711013}}].

\bibitem{Nekrasov:2010ka}
N.~Nekrasov and E.~Witten, {\it {The Omega Deformation, Branes, Integrability,
  and Liouville Theory}},  {\em JHEP} {\bf 09} (2010) 092,
  [\href{http://arxiv.org/abs/1002.0888}{{\tt arXiv:1002.0888}}].

\bibitem{Rapcak:2018nsl}
M.~Rapcak, Y.~Soibelman, Y.~Yang, and G.~Zhao, {\it {Cohomological Hall
  algebras, vertex algebras and instantons}},
  \href{http://arxiv.org/abs/1810.10402}{{\tt arXiv:1810.10402}}.

\bibitem{Chuang:2019qdz}
W.-y. Chuang, T.~Creutzig, D.~E. Diaconescu, and Y.~Soibelman, {\it {Hilbert
  schemes of nonreduced divisors in Calabi-Yau threefolds and W-algebras}},
  \href{http://arxiv.org/abs/1907.13005}{{\tt arXiv:1907.13005}}.

\bibitem{Koroteev:2019byp}
P.~Koroteev, {\it {On Quiver W-algebras and Defects from Gauge Origami}},
  \href{http://arxiv.org/abs/1908.04394}{{\tt arXiv:1908.04394}}.

\bibitem{Kontsevich:2010px}
M.~Kontsevich and Y.~Soibelman, {\it {Cohomological Hall algebra, exponential
  Hodge structures and motivic Donaldson-Thomas invariants}},  {\em Commun.
  Num. Theor. Phys.} {\bf 5} (2011) 231--352,
  [\href{http://arxiv.org/abs/1006.2706}{{\tt arXiv:1006.2706}}].

\bibitem{Nekrasov:2016qym}
N.~Nekrasov, {\it {BPS/CFT correspondence II: Instantons at crossroads, moduli
  and compactness theorem}},  {\em Adv. Theor. Math. Phys.} {\bf 21} (2017)
  503--583, [\href{http://arxiv.org/abs/1608.07272}{{\tt arXiv:1608.07272}}].

\bibitem{Nekrasov:2016gud}
N.~Nekrasov and N.~S. Prabhakar, {\it {Spiked Instantons from Intersecting
  D-branes}},  {\em Nucl. Phys.} {\bf B914} (2017) 257--300,
  [\href{http://arxiv.org/abs/1611.03478}{{\tt arXiv:1611.03478}}].

\bibitem{Prochazka:2018tlo}
T.~Procházka and M.~Rapčák, {\it {$ \mathcal{W} $ -algebra modules, free
  fields, and Gukov-Witten defects}},  {\em JHEP} {\bf 05} (2019) 159,
  [\href{http://arxiv.org/abs/1808.08837}{{\tt arXiv:1808.08837}}].

\bibitem{Negut:2017hxr}
A.~Neguţ, {\it {AGT relations for sheaves on surfaces}},
  \href{http://arxiv.org/abs/1711.00390}{{\tt arXiv:1711.00390}}.

\bibitem{Dedushenko:2017tdw}
M.~Dedushenko, S.~Gukov, and P.~Putrov, {\it {Vertex algebras and 4-manifold
  invariants}},  in {\em {Proceedings, Nigel Hitchin's 70th Birthday Conference
  : Geometry and Physics : A Festschrift in honour of Nigel Hitchin : 2
  volumes: Aarhus, Denmark, Oxford, UK, Madrid, Spain, September 5-16, 2016}},
  vol.~1, pp.~249--318, 2018.
\newblock \href{http://arxiv.org/abs/1705.01645}{{\tt arXiv:1705.01645}}.

\bibitem{Feigin:2018bkf}
B.~Feigin and S.~Gukov, {\it {VOA[$M_4$]}},
  \href{http://arxiv.org/abs/1806.02470}{{\tt arXiv:1806.02470}}.

\bibitem{Rapcak:2019abg}
M.~Rapcak, {\em {The Vertex Algebra Vertex}}.
\newblock PhD thesis, U. Waterloo (main), 2019.

\bibitem{Aganagic:2005wn}
M.~Aganagic, D.~Jafferis, and N.~Saulina, {\it {Branes, black holes and
  topological strings on toric Calabi-Yau manifolds}},  {\em JHEP} {\bf 12}
  (2006) 018, [\href{http://arxiv.org/abs/hep-th/0512245}{{\tt
  hep-th/0512245}}].

\bibitem{Jafferis:2006ny}
D.~Jafferis, {\it {Crystals and intersecting branes}},
  \href{http://arxiv.org/abs/hep-th/0607032}{{\tt hep-th/0607032}}.

\bibitem{Aganagic:2012si}
M.~Aganagic and K.~Schaeffer, {\it {Refined Black Hole Ensembles and
  Topological Strings}},  {\em JHEP} {\bf 01} (2013) 060,
  [\href{http://arxiv.org/abs/1210.1865}{{\tt arXiv:1210.1865}}].

\bibitem{tomas}
L.~Eberhardt and T.~Proch\'{a}zka, {\it {The matrix-extended
  $\mathcal{W}_{1+\infty}$ algebra, to appear}}, .

\bibitem{Creutzig:2018pts}
T.~Creutzig and Y.~Hikida, {\it {Rectangular W-algebras, extended higher spin
  gravity and dual coset CFTs}},  {\em JHEP} {\bf 02} (2019) 147,
  [\href{http://arxiv.org/abs/1812.07149}{{\tt arXiv:1812.07149}}].

\bibitem{Aganagic:2003db}
M.~Aganagic, A.~Klemm, M.~Marino, and C.~Vafa, {\it {The Topological vertex}},
  {\em Commun. Math. Phys.} {\bf 254} (2005) 425--478,
  [\href{http://arxiv.org/abs/hep-th/0305132}{{\tt hep-th/0305132}}].

\bibitem{Genra:2019tgw}
A.~Linshaw and F.~Malikov, {\it {One example of a chiral Lie group}},
  \href{http://arxiv.org/abs/1902.07414}{{\tt arXiv:1902.07414}}.

\bibitem{Prochazka:2019dvu}
T.~Procházka, {\it {Instanton R-matrix and W-symmetry}},
  \href{http://arxiv.org/abs/1903.10372}{{\tt arXiv:1903.10372}}.

\bibitem{Costello:2016nkh}
K.~Costello, {\it {M-theory in the Omega-background and 5-dimensional
  non-commutative gauge theory}},  \href{http://arxiv.org/abs/1610.04144}{{\tt
  arXiv:1610.04144}}.

\bibitem{Creutzig:2019qos}
T.~Creutzig and Y.~Hikida, {\it {Rectangular W-(super)algebras and their
  representations}},  \href{http://arxiv.org/abs/1906.05868}{{\tt
  arXiv:1906.05868}}.

\bibitem{Feigin:2013fga}
B.~Feigin, M.~Jimbo, T.~Miwa, and E.~Mukhin, {\it {Branching rules for quantum
  toroidal gl$_n$}},  {\em Adv. Math.} {\bf 300} (2016) 229--274,
  [\href{http://arxiv.org/abs/1309.2147}{{\tt arXiv:1309.2147}}].

\bibitem{Bershtein:2019shb}
M.~Bershtein and A.~Tsymbaliuk, {\it {Homomorphisms between different quantum
  toroidal and affine Yangian algebras}},  {\em J. Pure Appl. Algebra} {\bf
  223} (2019) 867--899.

\bibitem{Gaberdiel:2017hcn}
M.~R. Gaberdiel, W.~Li, C.~Peng, and H.~Zhang, {\it {The supersymmetric affine
  Yangian}},  {\em JHEP} {\bf 05} (2018) 200,
  [\href{http://arxiv.org/abs/1711.07449}{{\tt arXiv:1711.07449}}].

\bibitem{Gaberdiel:2018nbs}
M.~R. Gaberdiel, W.~Li, and C.~Peng, {\it {Twin-plane-partitions and
  $\mathcal{N}=2$ affine Yangian}},  {\em JHEP} {\bf 11} (2018) 192,
  [\href{http://arxiv.org/abs/1807.11304}{{\tt arXiv:1807.11304}}].

\bibitem{Li:2019nna}
W.~Li and P.~Longhi, {\it {Gluing two affine Yangians of $\mathfrak{gl}_1$}},
  \href{http://arxiv.org/abs/1905.03076}{{\tt arXiv:1905.03076}}.

\bibitem{Witten:2011zz}
E.~Witten, {\it {Fivebranes and Knots}},
  \href{http://arxiv.org/abs/1101.3216}{{\tt arXiv:1101.3216}}.

\bibitem{Mikhaylov:2014aoa}
V.~Mikhaylov and E.~Witten, {\it {Branes And Supergroups}},  {\em Commun. Math.
  Phys.} {\bf 340} (2015), no.~2 699--832,
  [\href{http://arxiv.org/abs/1410.1175}{{\tt arXiv:1410.1175}}].

\bibitem{Gaiotto:2008sa}
D.~Gaiotto and E.~Witten, {\it {Supersymmetric Boundary Conditions in N=4 Super
  Yang-Mills Theory}},  {\em J. Statist. Phys.} {\bf 135} (2009) 789--855,
  [\href{http://arxiv.org/abs/0804.2902}{{\tt arXiv:0804.2902}}].

\bibitem{Wakimoto:1986gf}
M.~Wakimoto, {\it {Fock representations of the affine lie algebra A1(1)}},
  {\em Commun. Math. Phys.} {\bf 104} (1986) 605--609.

\bibitem{Feigin:1990qn}
B.~L. Feigin and E.~V. Frenkel, {\it {Representations of affine Kac-Moody
  algebras, bosonization and resolutions}},  {\em Lett. Math. Phys.} {\bf 19}
  (1990) 307--317.

\bibitem{Yang:2007zzb}
W.-L. Yang, Y.-Z. Zhang, and X.~Liu, {\it {Free field realization of current
  superalgebra $gl(m|n)(k)$}},  {\em J. Math. Phys.} {\bf 48} (2007) 053514,
  [\href{http://arxiv.org/abs/0806.0190}{{\tt arXiv:0806.0190}}].

\end{thebibliography}\endgroup
\bibliographystyle{JHEP}

\end{document}